\documentclass{aa}
\usepackage{graphicx}
\usepackage{txfonts}
\usepackage{hyperref,textcomp}
\usepackage{ulem}
\hypersetup{
    colorlinks=true,
    linkcolor=red,
    urlcolor=magenta,
}

\usepackage[dvipsnames]{xcolor}

\newcommand{\cds}[1]{{\textcolor{blue} {#1}}}

\begin{document}
   \title{Shape, alignment, and mass distribution of baryonic and dark-matter halos in one EAGLE simulation
}
\author{
Q. Petit\inst{1},
C. Ducourant\inst{1},
E. Slezak\inst{2},
D. Sluse\inst{3},
L. Delchambre\inst{3}
}
  \institute{
          \inst{1} Laboratoire d'Astrophysique de Bordeaux, Univ. Bordeaux, CNRS, B18N, all{\'e}e Geoffroy Saint-Hilaire, 33615 Pessac, France\\
          \inst{2} Université Côte d'Azur, Observatoire de la Côte d'Azur, CNRS, Laboratoire Lagrange, Boulevard de l'Observatoire, CS 34229, 06304 Nice, France\\
          \inst{3} Institut d'Astrophysique et de Géophysique, Université de Liège, 19c, Allée du 6 Août, B-4000 Liège, Belgium\\
          \email{quentin.petit.1@u-bordeaux.fr}}

   \date{Received September 8, 2022 ; accepted December 5, 2022}

  \abstract
   {Accurate knowledge of the morphology of halos and its evolution are key constraints on the galaxy formation model as well as a determinant parameter of the strong-lensing phenomenon. Large-scale cosmic simulations are a tailored tool used to obtain statistics on the shape and mass distributions of these halos according to redshift.}
   {Using the cosmological hydrodynamic simulation, the Evolution and Assembly of GaLaxies and their Environments (EAGLE), we aim to provide a comprehensive analysis of the evolution of the morphology of galaxy halos and of their mass distributions with a focus on the snapshot at redshift $z=0.5$.}
   {We developed an iterative strategy involving a principal component analysis (PCA) to investigate the properties of the EAGLE halos and the differences in alignment between the various components. The semi-axes and orientation of the halos are estimated taking into account sub-halos in the simulation. The mass distributions of the dark-matter (DM), gas, and star halos are characterised by a half-mass radius, a concentration parameter and (projected) axis ratios.}
   {We present statistics of the shape parameters of 336\,540 halos from the EAGLE RefL0025N0376 simulation and describe their evolution from redshift $z=15$ to $z=0$. We measured the three-dimensional shape parameters for the DM, the gas, and the star components as well as for all particles. We also measured these parameters for two-dimensional projected distributions. At $z=0.5$, the minor axis of gas aligns with the minor axis of DM for massive halos ($M>10^{12}$ M$_\sun$), but this alignment is poorer for less massive halos. The DM halos axis ratios $b/a$ and $c/a$ have median values of $0.82 \pm 0.11$ and $0.64 \pm 0.12$, respectively. The gas in halos that also contain stars has a more flattened shape, with $b/a=0.70 \pm 0.19$ and $c/a = 0.38 \pm 0.20$. The sphericity of gas in halos w/ and w/o stars appears to be negatively correlated to the total mass, while the sphericity of DM is insensitive to it. The measured projected axis ratios, $b_p/a_p$, of star halos at $z=0.5$ have a median value of $0.80 \pm 0.07$, which is in good agreement with ground-based and space-based measurements within 1 $\sigma$. For DM halos, we measure a value of $0.85 \pm 0.06$. The evolution of the concentration as a function of redshift is fairly homogeneous for the various components, except for the starless gas halos, which appear much more concentrated for $z>0.7$.}
{}
   


\titlerunning{Shape and alignment of baryonic and DM halos}
\authorrunning{Q. Petit et al.}
\maketitle
\section{Introduction}\label{intro}

The widely accepted $\rm \Lambda$-cold dark matter ($\rm \Lambda$-CDM) model postulates that a highly non-uniform system of voids, sheets, filaments, and halos — commonly known as the cosmic web — emerges as a result of the gravitational growth of tiny heterogeneities in the early Universe's matter distribution \citep{1996Natur,2008Sci}. The evolution over time of this large-scale cosmic web is regulated by the dark-energy (DE) and dark-matter (DM) components, since they are today by far the main contributors to the Universe's total mass-energy content. Regarding the visible galaxies, which gather most of the baryonic content, a key assumption of galaxy formation models is that these galaxies develop within the DM halos that permeate the cosmic web \citep{1978white,1984Natur}. However, baryonic structures do not always directly trace the form and shape of their DM halos due to the wide variety of dissipative physical processes that are involved in the evolution of galaxies. Through the redistribution of angular momentum, the geometry of the DM halo influences certain properties of galaxies, such as the development and evolution of bars \citep{2018Collier,2019Collier,2022Kumar}.

There are other observational influences on large-scale mass distribution such as weak and strong lensing. With the emergence of weak gravitational lensing as a precision probe of cosmology, galaxy alignments have taken on an added importance because they can mimic cosmic shear, such as the effect of gravitational lensing by a large-scale structure on observed galaxy shapes \citep{2015Kirk}. The shape parameters of the mass distribution of foreground galaxies can be measured from the two-dimensional shear field derived from background galaxies \citep{2000Natarajan}. Indeed, the distribution of satellite galaxies preferentially aligns with the major axis of the central galaxy, with a clear dependence on both halo mass and galaxy colours \citep{2008Wang}. Another important tool to probe the total matter distribution of galaxies is strong gravitational lensing\footnote{Strong gravitational lensing depicts the formation of multiple images of a background source whose light rays are deflected due to the presence of a large mass concentration like a massive galaxy or a galaxy cluster in the line of sight between the observer and the source.}. In that context, quadruply imaged quasars (quads) are particularly interesting because of the constraints they put on the radial and azimutal mass distribution within the deflectors. In multiple imaged lens systems, the observed configuration of images is completely dependent on the projected mass distribution of the deflecting galaxy. Lenses are sought in many large and deep sky surveys such as the Sloan Digital Sky Survey \citep{2004SDSS}, the Dark Energy Survey \citep{2016DES} and the Gaia data \citep{2018Krone,2019Delchambre,2021Stern} with machine-learning algorithms. Known lenses are too rare to be used for the training of these algorithms. One needs a set of realistic simulations for that with in particular a realistic distribution of projected ellipticities of the deflecting galaxies. Recent studies show that using a non-singular isothermal ellipsoid projected mass distribution with the addition of an external shear as a lens model helps reproduce the distribution of observed configurations \citep{1996Witt,2015Woldesenbet,2021Gomer}. Ongoing searches (such as those performed by Gaia GraL) allow one to establish samples with well understood selection functions, and pave the way to upcoming facilities such as Euclid, which will increase the number of known lenses by a few orders of magnitude.

Observational evidence for non-spherical DM halos \citep{2012Barnabe,2012vanUitert} requires an account of baryonic physics on galaxy formation. Running cosmological numerical simulations is a way to improve our understanding of the building-up of galaxies by studying how simulated galaxies manifest themselves in a variety of morphologies ranging from thin to bulge-dominated discs and to ellipsoids of various sizes and masses. To simulate galaxies with statistical characteristic properties similar to the ones measured in real catalogues, it is crucial to combine the evolution of large-scale DM halos with the study of baryon physics at the smallest scales. Hence, highly resolved simulations large enough to include the large-scale structure while capturing the morphology and underlying physics of galaxies are required. Progress has been made in this regard by the current generation of hydrodynamical simulations such as the Magneticum Pathfinder \citep{2015magneticum}, the MassiveBlack-II \citep{2015Khandai}, the EAGLE project \citep{eagleproject}, and more recently the IllustrisTNG simulation \citep{2019Illustris50}. With these numerical simulations, it is possible to make detailed predictions of the cosmological distribution and evolution of the galaxies that can be compared with a wide range of observations, including future large-scale structure surveys, such as those that will be produced by the Vera Rubin Observatory \citep{2009lsst}, the Euclid spacecraft \citep{euclid}, and the Nancy Grace Roman Space Telescope \citep{2015nancy}.

The EAGLE suite of simulations is particularly well suited to this application because these simulations were explicitly calibrated to reproduce the $z=0$ observed galaxy stellar mass function with realistic constrains on sizes. In this study, we used the hydrodynamic simulation with a volume of side $L=25$ cMpc, with $N=376^3$ initial DM particles \citep{eagleproject}. This simulation has the smallest volume but has the same resolution as larger volume simulations. Several groups have analyse various aspect of the morphologies of the galaxies in the EAGLE simulations \citep{2015Velliscig,2016Shao,hill2021}. In this paper we intend to complement these works on the morphology of DM, gas, and star halos by bringing in a new method to determine the shapes parameters of the halos and their orientations. We also analyse in details the mass distribution of the various components of the halos. 

We measure directly the shapes of the DM, gas and stars components of halos and subhalos (modeled as ellipsoids in three dimensional space). We examine how shapes evolve with time and as a function of halo total mass. We also measure the projected (2D) shapes for comparison with observations. By measuring projected ellipticities of the DM, gas and stars components of simulated galaxies, we can estimate the distribution of ellipticities. The shape of the halos together with statistics on the ellipticities will allow us to generate simulations of gravitational lenses that will be used to train an automatic detection tools for these objects. 

This paper is structured as follows. In Sect. \ref{data}, we briefly describe the EAGLE simulation with a focus on the content of the simulation at redsfhit $z=0.5$, which is the typical redshift of known gravitational lense deflectors. In Sect. \ref{method}, we present our methods to measure the shapes of the DM, gas, and star components of halos and sub-halos. In Sect. \ref{validation}, we validate our method on simulations and we study the impact of the number of particles and the axis ratios of the halos on the measured parameters. In Sect. \ref{red5}, we provide a comprehensive description of the shapes and the internal alignment of gas with DM and stars as a function of halo total mass at  redshift $z=0.5$. We also analyse the properties of the projected halos on the sky. In Sect. \ref{dependredshift}, we examine the dependence of the DM, gas, and star halos on the redshift and the halos' total masses. In Sect. \ref{Ccl}, we discuss and summarise our findings.

\section{Data}\label{data}
\subsection{The EAGLE simulations}\label{eagledata}

The Evolution and Assembly of GaLaxies and their Environments (EAGLE) project \citep{eagleproject} is a suite of cosmological, hydrodynamic simulations aimed at studying the evolution of DM and gas in the Universe and the formation of stars and black holes. These simulations are computed using a modified version of the GADGET-3 tree-SPH code from \citep{2005springel}. They include sub-grid descriptions for radiative cooling \citep{2009aWiersma}, star formation \citep{2008starformation}, multi-element metal enrichment \citep{2009bWiersma}, black hole formation \citep{2005SpringelBlackHole,2015Rosas}, as well as feedback from massive stars and active galactic nucleus (AGNs)  \citep{2012Dalla,eagleproject}. The EAGLE public database, which includes the EAGLE galaxy database \citep{alpine2016} and EAGLE particle data \citep{eagleteam2017}, can be accessed and queried via their website\footnote{\url{http://icc.dur.ac.uk/Eagle/}}.

Each simulation includes 29 snapshots at different redshifts between $z=0$ and $z=15$ allowing us to study the evolution of the morphology of galaxies along time. Our primary analysis focuses on the EAGLE simulation with the smallest cosmic volume, RefL0025N0376. This choice is motivated by the need to extensively test our analysis code while limiting the calculation time. This simulation follows a cubic periodic volume of side $L=25$ cMpc, with $N=376^3$ collision-less DM particles with an individual mass of $M_{\rm \rm DM}=9.7 \times 10^6$ M$_\sun$ and a starting equal number of baryonic (gas) particles, each with an initial mass of $m_g=1.81 \times 10^6$ M$_{\sun}$. The initial conditions were created using values of the cosmological parameters derived from the initial Planck data release \citep{planck2014}, namely $\Omega_0=0.307$, $\Omega_b\simeq0.048$, $\Omega_\Lambda=0.693$, $\sigma_8\simeq0.83$, $n_s=0.9611$, and $h\simeq0.68$, $Y=0.248$. The ratio of the baryonic-to-non-baryonic particles is therefore equal to $(\Omega_0 - \Omega_b)/\Omega_b \equiv (m_{DM}/m_{b}) \simeq 5.4$.

\begin{figure}
    \resizebox{\hsize}{!}{\includegraphics{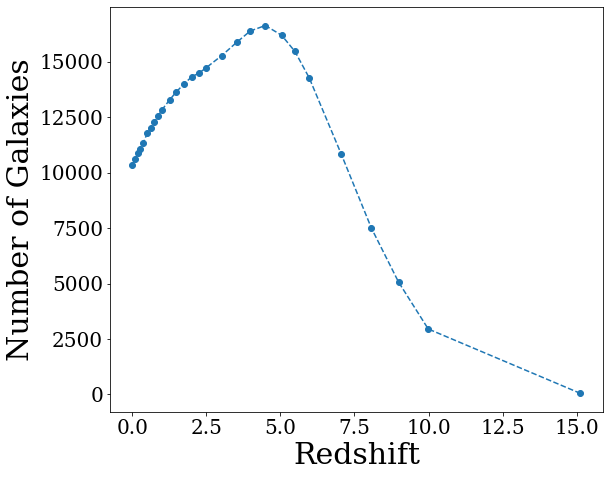}}
  	\caption{Number of halos studied in this paper as a function of redshift in the RefL0025N0376 EAGLE simulation.}
    \label{number_galaxies}
\end{figure}

\subsection{Star formation and its energy feedback}

In the EAGLE simulation, stars are formed from gas particles when the gas density exceeds a given threshold \citep[see][for a detailed presentation]{eagleproject}. The limited resolution of the simulation makes it necessary to impose such a threshold for star formation for a density above which a cold interstellar gas phase is expected to form.
The probability that a gas particle is converted into a collision-less star particle depends on the gas particle's pressure \citep{2004schaye,schayeproba} and metallicity. The evolution of stars is tracked using the metallicity-dependent lifetimes of \cite{1998Portinari}. The mass lost by stars is distributed among the neighbouring particles through winds from B-, A-, and G-type stars, winds from massive stars, and core-collapse supernovae using the nucleosynthetic yields from \cite{2001Marigo}. Stars can also inject energy and moment into the interstellar medium (ISM) through stellar winds, radiation, and supernovae. These processes are particularly important for massive stars. 

\subsection{Black holes and feedback from AGN}

Supermassive black holes (BHs) are settled in halos with a total mass greater than $10^{10}$ M$_\sun$ and which do not already contain a black hole \citep{2005springel,2009booth,eagleproject}. These halos are identified by a friends-of-friends (FoF) algorithm at epochs corresponding to a logarithmic sampling of the expansion factor. The gas particle whose density is the highest is converted into a collision-less BH particle with sub-grid BH mass $m_{BH}=10^5$ M$_\sun/h$. BHs grow by accretion of gas. This accretion is dependent on the mass of the BH, the local density and temperature, the velocity of the BH relative to the ambient gas, and the angular momentum of the gas with respect to the BH. BHs are merged with other BHs if they are separated by a distance that is smaller than both the smoothing kernel of the BH, $h_{BH}$, and three gravitational softening lengths. A limit on the allowed relative velocity prevents BHs from merging during the initial stages of halo mergers.

The feedback associated with AGNs is implemented in a manner analogous to energy feedback from star formation. This feedback is tuned so that the simulation reproduces the present-day galaxy stellar mass function and realistic sizes for disc galaxies \citep{2015Crain}, and it verifies the close relationship between the stellar mass of galaxies and the mass of their central BHs.

\subsection{Halo identification and sample}

As mentioned, DM structures are initially identified using the FoF algorithm with a linking length of 0.2 times the mean inter-particle spacing \citep{1985Davis}. Particles other than the DM ones (gas, stars, and black holes) are assigned to the same group as their nearest DM neighbours. Halos are considered as gravitationally self-bound structures within the FoF groups using the SUBFIND algorithm \citep{2001springel,2009dolag}. Unlike FoF, SUBFIND considers all types of particle and identifies self-bound sub-units within a bound structure that are called sub-halos. The local minima in the gravitational potential field are the centres of sub-halo candidates. The sub-halo with a minimum value of the gravitational potential within a FoF group is defined as the 'main halo'. Any particle bound to the group but not assigned to any other sub-halos is assigned to the main halo. In this study, sub-halos are defined as satellite halos to the main halo. 

Our study is based on the evolution of the smallest EAGLE simulation, RefL025N0376, spanning a stellar mass range from $10^{9}$ to $10^{11}$ M$_\sun$ over a redshift range of $0 \leq z \leq 15$. Fig. \ref{number_galaxies} presents the number of halos from the small simulation according to the snapshot redshifts. The number of detected halos first increases from $z=15$ to $z=5$ as their difference in local density is getting larger due to their gravitational build-up. Then, at redshift $z \approx 5$, the hierarchical formation of large and massive halos becomes dominant, leading to a decrease in the total number of individual structures.

\subsection{Overall properties of the halos at $z=0.5$}\label{snapshot23}

In this section, we present the halo content of the RefL0025N0376 simulation, and for that we chose the redshift $z=0.5$, which is the typical redshift of known gravitational lense deflectors. We analyse the mass distribution and look at the distribution of the different types of particles in the halos and sub-halos. 


The mass distribution of the DM particles remains constant in the simulation ($m_{DM}=9.7 \times 10^6\,$M$_\sun$). DM particles are non-colliding and therefore only interact gravitationally. Several gas particles are more massive than their initial mass ($m_g=1.81 \times 10^6\,$M$_\sun$) but this situation remains rare due to the high probability of these particles being converted into star particles. As already mentioned, the initial mass ratio between DM and gas particles is exactly the ratio of the density of these two components in the cosmological model used by the EAGLE simulation.

\begin{figure}
    \resizebox{\hsize}{!}{\includegraphics{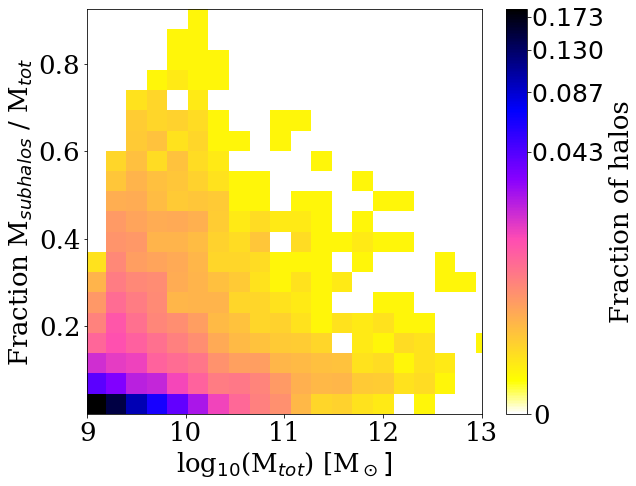}}
  	\caption{Fraction of the sub-halo mass to the total mass of the halo as a function of the total mass of halos. Colours represent the fraction of halos at redshift z=0.5.}
    \label{ratio_sub_main}
\end{figure}

In Fig. \ref{ratio_sub_main}, we present the fraction of the sub-halo mass to the total mass of the halo as a function of the total mass of halos. In this way, we can detect the presence of massive sub-halos in comparison to their main parent halo. Most halos have a mass between $10^9$ and $10^{11}$ M$_{\sun}$. For the less massive halos, a large proportion of the mass can be accounted for by the sub-halos. This is an issue when analysing the morphology of such halos. To take it into account, any halo whose sub-halos contribute to more than 10\% of its total mass are not considered in the results. We found that 88\% of all halos identified at redshift $z=0.5$ have less than 10\% of their total mass in sub-halos. 

\begin{figure}
    \resizebox{\hsize}{!}{\includegraphics{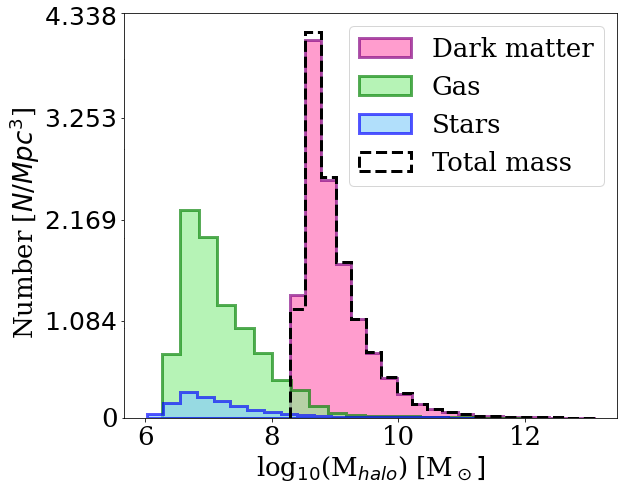}}
  	\caption{Mass distribution of 42 766 EAGLE halos at redshift $z=0.5$ for each type of particle. The y-axis shows the volume densities for a box size of 16.94 Mpc per side. The black dotted line represents the total mass of the halos by summing the mass of all the particles. We notice that the DM particles dominate the halos masses by several orders of magnitude.}
    \label{massdistribution}
\end{figure}

Figure \ref{massdistribution} presents the distribution of the mass for the 42 766 halos considered. The DM mass contribution is several orders of magnitude larger than those of the two other types of particles. As expected, the mass of the halos is largely dominated by DM.

\begin{figure}
    \resizebox{\hsize}{!}{\includegraphics{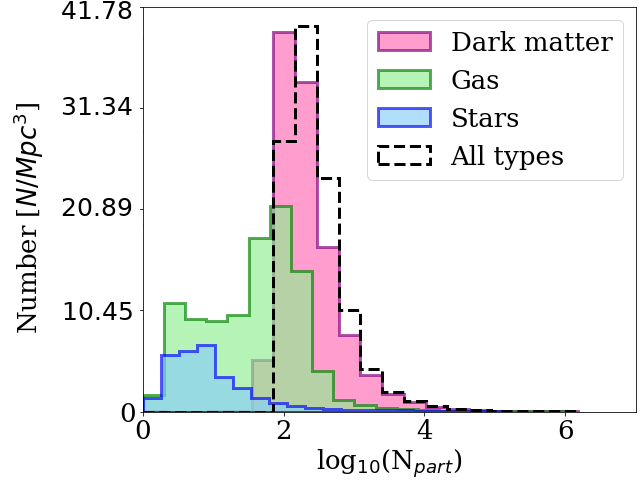}}
  	\caption{Distribution of number of particles of each type in halos in RefL0025N0376 simulation at redshift $z=0.5$. The y-axis shows the volume densities for a box size of 16.94 Mpc per side.}
    \label{number_particles}
\end{figure}

Figure \ref{number_particles} presents the number of particles in these halos for each type of particle. There are very few massive black holes in each halo, so they do not directly influence the overall shape. Therefore, they will only be accounted for in the total mass analysis of the halos. On average, 50\% of the halos have a small number of DM particles (50-300 particles), but the number of baryonic particles is even smaller (only 7\% have more than 50 gas particles and 2\% more than 50 stars). As a result, we observe many dark halos containing very few or no baryonic particles (90\% of halos have no stars). 

\section{Shape analysis}\label{method}

The EAGLE halos have various shapes and sizes and often have rich structures. In order to analyse all the simulations and model each halo whatever its shape is, we developed a specific analysis procedure that is detailed hereafter. It was applied five times, respectively, for each type of particle (DM, gas, stars) as well as for the sum of the baryonic components and all particles together.

We also performed a complementary shape analysis of the projected halos in the [xy], [yz], and [xz] plans to derive the statistical distribution of the shape parameters of once projected on a plane that is tangential to the celestial sphere. The projected mass distribution being a central parameter of the modelling of gravitational lenses.

\begin{figure}
    \resizebox{\hsize}{!}{\includegraphics{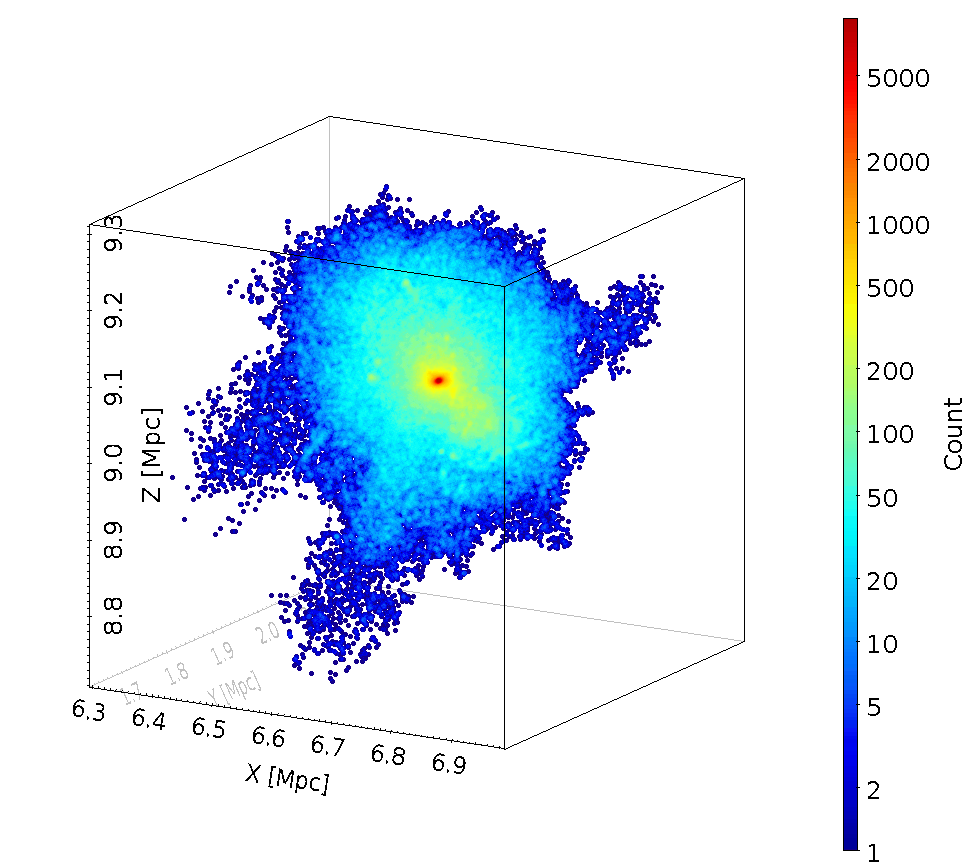}}
  	\caption{Three dimensional plot of main halo \#20 from the RefL0025N0376 simulation at z=0.5 with a large-scale sub-structure in the down part of size $\sim150$ kpc. Coordinates are in the initial EAGLE simulation reference system (X,Y,Z).  All types of particles are represented.}
    \label{3D}
\end{figure}

An accurate measurement of the major and minor axes of an EAGLE halo is not straightforward, because generally the distribution of particles is not homogeneous and presents peripheral sub-structure, as illustrated in Fig. \ref{3D}, that impacts the measurements. Recent studies of the EAGLE simulations focus on their star and gas content \citep{thob2019,hill2021} and only study the central part of the halos. The present work intends to assess the overall morphology at a larger scale accounting for most of the matter including DM, which is the gravitational dominant component. A specific strategy was, however, developed to get rid of halo sub-structures.

\subsection{Elimination of peripheral sub-structures for shape analysis}
\subsubsection{Determination of the centre of the halo}

We started by determining the centre of the main halo containing all the particles without taking into account the influence of peripheral sub-structures. First, we considered all particles and found their median position (x0,y0,z0). Then, a new median position (x0',y0',z0') was derived from the particles encompassed in a sphere centred on (x0,y0,z0) and with a radius equal to one tenth of the maximal distance between any particle and the centre, giving priority to the densest part of the halo. This second step is re-run but using the new centre (x0',y0',z0') until convergence is reached and the final median position of the remaining particles defines the final centre of the halo.

A visual check is used to monitor convergence. This centre corresponds to the location of the highest density peak of particles. In practice, three iterations are often required before convergence. 

\subsubsection{Working sphere}

In order to confidently measure the shape properties of a halo, we need to get rid of the influence of peripheral sub-structures. For this, we aim to determine a working spherical volume characterised by its radius, $r_{\rm 80cc}$, that encompasses a sufficient amount of mass while not including the perturbing sub-structures. $r_{\rm 80cc}$ is empirically chosen as the radius at which the density drops by 80\% from the central core density. The computed sphere then delimits the volume comprising the particles we chose to consider.

To derive $r_{\rm 80cc}$, it is necessary to establish the density profile of the halo. For this, we first need to determine the size of the spherical core of the halo,  $r_{\rm core}$, that contains 10\% of the total mass and within which the core density will be measured.

\begin{figure}
    \resizebox{\hsize}{!}{\includegraphics{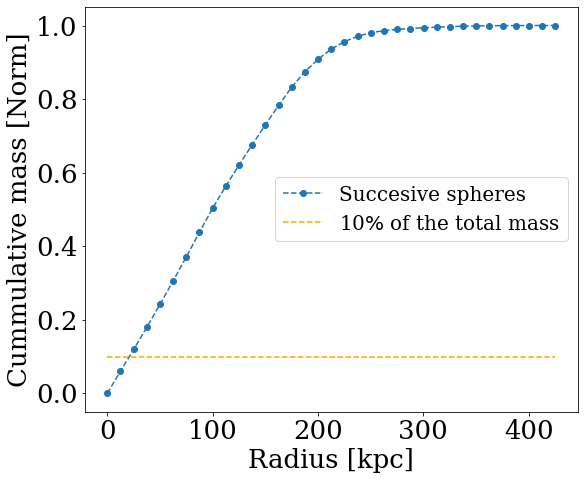}}
  	\caption{Cumulative mass growth curve of the main halo \#20, obtained from mass measurement in successive concentric spherical shells with width of 1/50 of the total spatial extension. The intersection of the yellow line with the growth curve indicates the core radius $r_{\rm core}$ encompassing 10\% of the total mass. Here $r_{\rm core}$=20.92.}
    \label{spheric_growth}
\end{figure}

To do so, we produce a cumulative mass growth curve of the halo (as shown in Fig. \ref{spheric_growth}) by summing the mass of particles in successive spherical shells with width equals to 1/50 of the total spatial extension of the halo. When the derived mass does not change from the mean of the last three measurements by more than $1 \sigma$, the process stops. The asymptotic value of the mass growth curve corresponds to the total mass of the halo. From this curve, we determine the core radius of the halo, encompassing 10\% of the total (asymptotic) mass.

\begin{figure}
    \resizebox{\hsize}{!}{\includegraphics{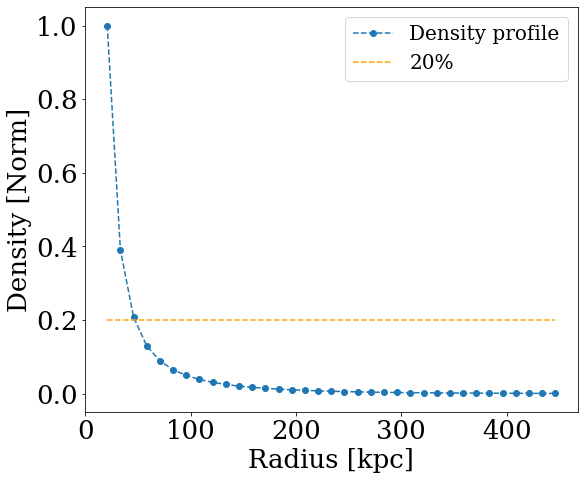}}
  	\caption{Density profile of main halo \#20, obtained from density estimation in successive concentric spherical shells with width of 1/50 of the total spatial extension. The intersection of the yellow line with the density curve indicates the radius $r_{\rm 80cc}$, at which point the density drops by 80\% from the central core density. Here $r_{\rm core}$=20.92 kpc and $r_{\rm 80cc}$=47.25 kpc.}
    \label{density}
\end{figure}

 It is then possible to establish the radial density profile of the halo by estimating the density in the core, $r_{\rm core}$, and in successive spherical shells with a radius step of 1/50 of the total spatial extension of the halo. We present in Fig. \ref{density} the resulting density profile.

The radius at which the density drops by 80\% from the central core density, $r_{\rm 80cc}$, is derived from this figure via a linear interpolation. This radius defines the sphere in which  the following shape analysis is performed.

\subsection{Semi-axis and shape parameter estimation}
The semi-axes (a, b, and c, from largest to smallest) of the halo are estimated by applying a principal components analysis (PCA) to the spatial coordinates of the particles contained in the working sphere. The PCA determines the principal axes of the distribution that best explain the (normal) variance in the coordinate space. The length of each principal axis is given by the square root of the eigenvalues of the covariance matrix of the particles coordinates. From a practical point of view, the semi-axes measure the standard deviations of the underlying three dimensional particle distribution.

\begin{figure}
    \resizebox{\hsize}{!}{\includegraphics{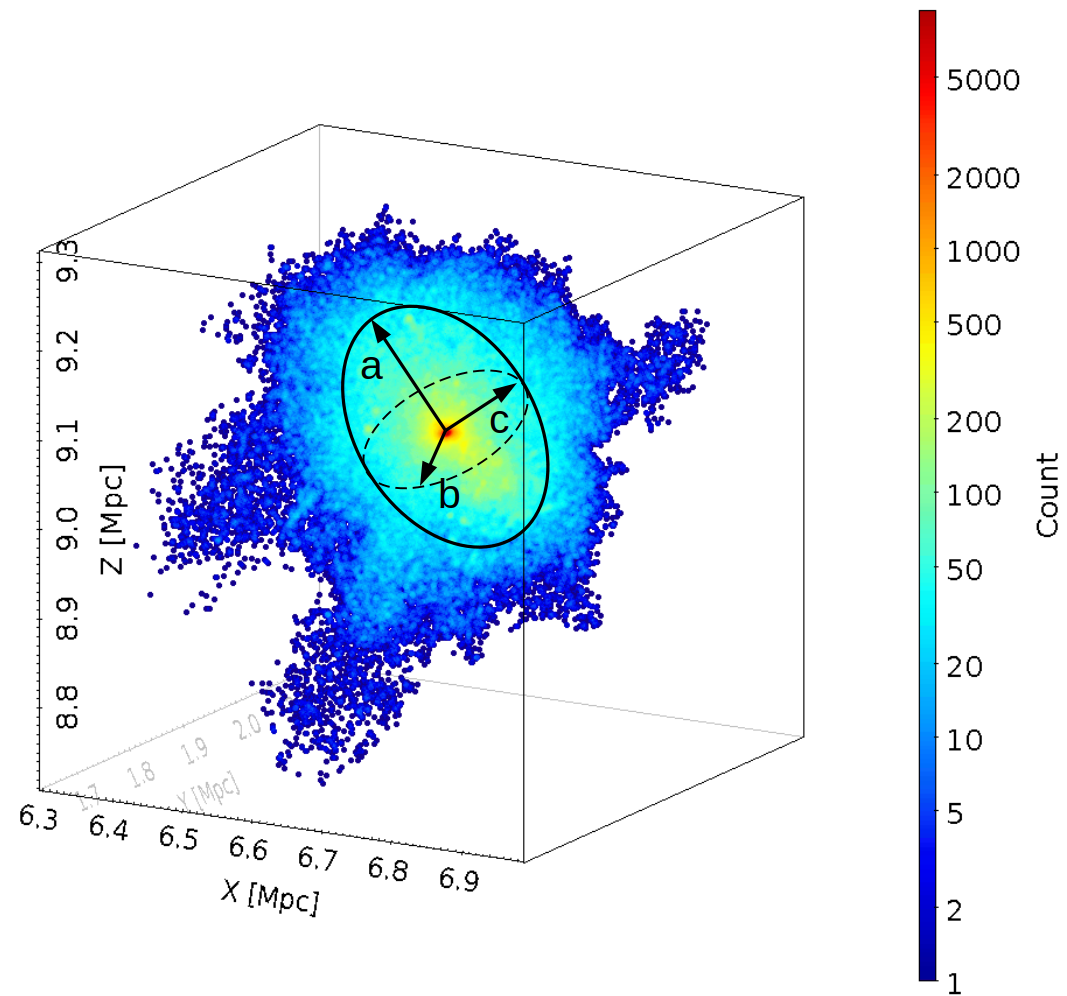}}
  	\caption{Three dimensional plot of main halo \#20 from EAGLE (z=0.5) with materialisation of the final ellipsoid describing the halo, obtained by an iterative PCA.}
    \label{3D_ellipse}
\end{figure}

Once the first estimation of the semi-axes is obtained, the PCA is re-applied iteratively to the selection of particles encompassed in the ellipsoid defined by the semi-axes derived at previous iteration and a new estimation of the semi-axes is produced. After a few iterations (<10) convergence is reached and the final semi-axis estimates (a,b,c) are obtained (see Fig. \ref{3D_ellipse}). By convention, $a > b > c$.

From the final estimates of the semi-axes of the fitted ellipsoid, we define the axis ratios $b/a$, $c/a$, and $c/b$. The ratio $c/a$ is also called sphericity ($c/a=1$ corresponds to a sphere, whereas $c/a=0$ corresponds to a flat disc if $b \neq 0$).

We have derived in this way, for each halo, estimates of the semi-axes  of the ellipsoids corresponding to the distribution of each type of particles (stars, gas, DM) as well as to the distribution of baryonic and all particles: (a, b, c)$_{star}$, (a, b, c)$_{gas}$, (a, b, c)$_{DM}$, (a, b, c)$_{baryonic}$, (a, b, c)$_{all}$.

\subsection{Orientation of halos}\label{orientation}
In the case of a three-dimensional distribution of particles, the principal axes of the distribution define a new orthogonal reference system (X',Y',Z'). This reference system is rotated with respect to the initial EAGLE reference system (X,Y,Z) by three Euler angles ($\psi, \theta, \phi$) that are given in the rotation matrix. This matrix allows us to convert the position of the particles given in the EAGLE simulation reference system (X,Y,Z) to the new reference system defined by the principal axes (X',Y',Z') of the distribution.

We derived estimates of the three Euler angle this way for each halo, for each type of particle, as well as for the entirety as ($\psi$, $\theta$, $\phi$)$_{stars}$, ($\psi$, $\theta$, $\phi$)$_{gas}$, ($\psi$, $\theta$, $\phi$)$_{DM}$, ($\psi$, $\theta$, $\phi$)$_{all}$.  

\subsection{Mass growth curve and mass concentration}\label{mass_g}
\begin{figure*}
\centering
    \includegraphics[width=9.6cm]{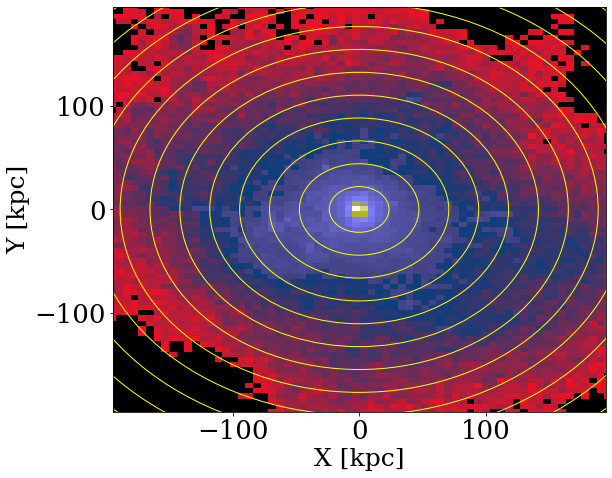}
    \includegraphics[width=8.6cm]{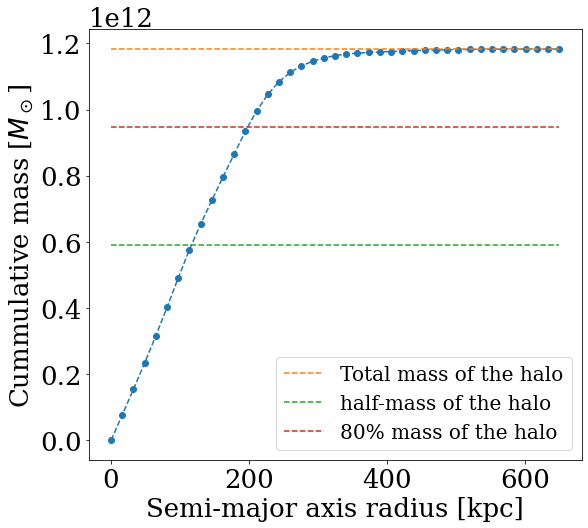}
    \caption{Calculation of the mass growth curve. (Left panel): Projection along (X,Y) of main halo \#20 of EAGLE redshift z=0.5. Yellow ellipses represent successive concentric ellipses. Here, only one in five ellipses are displayed. (Right panel): The mass of the particles contained in each successive ellipse is summed to construct the cumulative halo mass curve. The green dashed line is the asymptotic value of the mass growth curve, which corresponds to the normalised total mass of the halo. The red and blue dashed lines correspond to the radii containing 20\% and 80\% of the total mass, respectively.}
    \label{growth}
\end{figure*}

Once the shape of a halo is known, it is possible to establish the final mass growth curve characterising its mass distribution. This curve is computed by adding the mass of the particles contained in successive concentric ellipsoid shells with a fixed width, as illustrated in Fig. \ref{growth} (left panel). The calculation is stopped when an asymptote is reached. The resulting curve is smoothed by a moving average with a three-point window. Fig. \ref{growth} (right panel) presents the final mass growth curve of the main halo \#20 from EAGLE ($M = 1.18 \times 10^{12}\,$M$_\sun$, $z=0.5$, $N_{part} = 1.62 \times 10^5$ particles).

In galaxy classification studies, light concentration is often used to characterise the light distribution of galaxies in view of discriminating between early-types (bulge dominated) and late-type objects. A similar quantity, the mass concentration, can be estimated as \citep{2003Conselice}:

\begin{equation}
    C = 5 \times \log{\Bigg(\frac{r_{80}}{r_{20}}\Bigg)},
    \label{Concentration}
\end{equation}

where $r_{20}$ and $r_{80}$ are the radii containing 20\% and 80\% of the total mass, respectively. These radii are measured from the mass growth curve as the intersection of the curve with the horizontal lines corresponding to 20\% and 80\% of the total mass.

We derived a mass growth curve and a mass concentration this way for each halo, for each type of particle, as well as for the whole halo.

\section{Validation of the algorithm}\label{validation}

\subsection{Simulation of halos}

We validated our method by applying it to $10\,000$ simulated halos with shape parameters ($a_{sim}$, $b_{sim}$, $c_{sim}$) randomly selected such that a>b>c and a<1. To simulate halos, we draw the particles coordinates from a multivariate normal distribution with mean $\vec{\mu} = [ \mu_x, \mu_y, \mu_z$ ] and covariance matrix 
$\Sigma = diag (a_{\rm sim}^2, b_{\rm sim}^2, c_{\rm sim}^2)$. 
We then rotated the whole distribution of particles by three randomly selected Euler angles ($\psi_{sim}$, $\theta_{sim}$ and $\phi_{sim}$) in order to simulate randomly orientated halos. The resulting simulated halo has particle coordinates expressed in the original (X,Y,Z) reference system. We do not consider the possible presence of substructures in the halos since our algorithm is built to minimise the influence of these substructures before applying the morphological analysis. 

\subsection{Estimation of the halo's parameters}
Two parameters may influence the correct recovery of the input parameters defining the shape and orientation of the simulated halos: the number of particles in the halo (too sparse distribution may be poorly recovered) and the sphericity of the halo (distribution close to a sphere may be harder to retrieve). 

\subsubsection{Dependence on the number of particles}\label{dependence_nb}
To explore the dependence on the number of particles, we simulated $10\,000$ synthetic halos containing 10 to 2000 uniformly drawn particles, as observed in EAGLE halos (see Fig. \ref{number_particles}) and with a fixed sphericity. We did that for two different sphericities: $c_{sim}/a_{sim}$ = 0.72 (rather spherical halos) in one case and $c_{sim}/a_{sim}$ = 0.25 (flattened-disc halos) in the other. We then applied our algorithm to these synthetic halos and derived estimates of the semi-axes (a,b,c) and the Euler angles ($\psi$, $\theta$, $\phi$). 

In Appendix \ref{appendixvalidation}, we present the diagrams presenting the relative errors\footnote{$\Delta x = (x - x_{\rm sim}) / x_{\rm sim}$} ($\Delta a$, $\Delta b$, $\Delta c$) and the difference between the estimated angles ($\psi$, $\theta$, $\phi$) and their true value ($\psi_{sim}$, $\theta_{sim}$, $\phi_{sim}$) as a function of the number of particles and the sphericity.

The semi-axes are generally well recovered to a level better than 10\%, and generally to one better than 5\% for halos with more than 100 particles, whatever the sphericity is. There is a visible degradation of the quality of the semi-axes when the number of particles is smaller than 100. The semi-axis 'c' then tends to be slightly underestimated by less than 5\% for halos with fewer than 100 particles. We also notice that the orientation of the halos is globally well recovered to better than 10 deg in all cases. The sphericity of the halos is influencing the quality of the recovered orientation as expected: for quasi-spherical halos with more than 100 particles, the errors are of 8-10 deg or fewer, while for flattened halos they are around 2-3 deg. 

\subsubsection{Dependence on the sphericity}\label{depend_orientation}
To explore the dependence on the sphericity of the halos, we have simulated $10\,000$ synthetic halos of 500 particles with randomly selected axis ratios such that $(b_{sim}/a_{sim})$<1 and $(c_{sim}/b_{sim})$<1. We analysed these halos with our algorithm and derived estimates of their semi-axes (a,b,c) and their Euler angles ($\psi$, $\theta$, $\phi$).

The relative errors ($\Delta a$, $\Delta b$, $\Delta c$) on the estimated semi-axes and the difference between the estimated angles ($\psi$, $\theta$, $\phi$) and their true value as a function of their sphericity is presented in Appendix \ref{appendixvalidation}. We observe that the semi-axes (a,b,c) are well recovered but a slight systematic effect is observed for halos with high sphericity ($c/a$>0.9): major-axis $a$ is overestimated by 0.2\%, semi-axis $b$ is underestimated by 0.3\%, while $c$ does not seem to be influenced by the sphericity of the halos. The orientation of the halos is also well recovered with relative errors not influenced by the sphericity of the halo, except for $\phi$ (the third rotation). The distribution of quartiles of $\Delta\phi$ indicates a severe degradation (errors up to 10 deg and more) for sphericities larger than 0.9 with a systematic over-estimation of $\phi$.  

We conclude that the estimate of the value of the principal semi-axes are very well recovered when the number of particles included in the halos are larger than 100. The semi-axes are generally recovered at a rate better than $5\%$. The Euler angles of the halos are well recovered to better than 10 deg for halos with a number of particles larger than 100, but their recovery is dependent on the sphericity of the halo, degrading for quasi-spherical halos with c/a>0.9.

\section{Application to the EAGLE snapshot at redshift z=0.5}\label{red5}

We first focus on a comprehensive description of the results obtained for the redshift $z=0.5$ in the RefL0025N0376 simulation, because one motivation of our work consists of studying strong lensing. We limit our analysis to halos containing at least 100 particles of each type in order to ensure the reliability of the estimated parameters, as seen in Section \ref{dependence_nb}. We derive shape parameters describing (i) the DM distribution for 17\,480 halos, (ii) the gas distribution for 2\,122 halos, and (iii) the star distribution for 682 halos. From the 2\,122 halos with a sufficient number of gas particles to be considered as a halo, 1\,440 do not contain the minimum of 100 stars we required, while 682 do (hereafter gas halos 'w/ stars', gas halos 'w/o stars'). 

\begin{figure*}[h]
\centering
    \includegraphics[width=8.6cm]{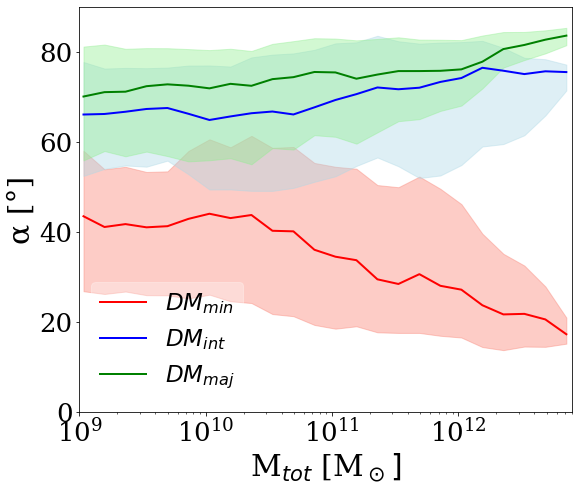}
    \includegraphics[width=8.6cm]{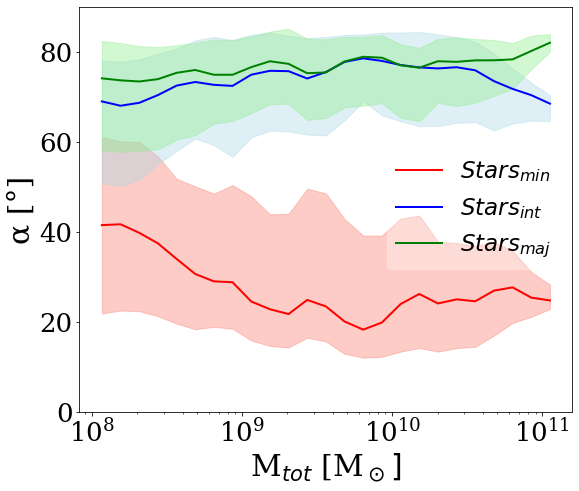}
  	\caption{Difference of alignment angle, $\alpha$, between the minor axis of gas halos and all semi-axis of the DM (left panel) and star halos (right panel). The solid lines indicate the median values of $\alpha$, whilst the shading denotes the 25th - 75th quartiles. The values are shown as a function of the total mass (M$_{tot}$). The red, blue, and green curves correspond to the difference of alignment of the gas minor axis with respect to the minor, intermediate, and major axes, respectively.} 
    \label{orienta}
\end{figure*}

\subsection{Orientation of the halos}\label{shapeparameters}

We begin with an examination of the differences in the alignment of the gas, stars and DM halos as a function of the total mass. We quantify the differences of alignment of the various components via the angle $\alpha$, between the semi-axes of the ellipsoid describing the distributions via the scalar product of two major (intermediate or minor) axes ($\alpha$ in case of projected halos). We note that it is more typical in the literature to use the major axis. In contrast, it is the minor axis that is the most natural choice when focusing on discs. The quantity $\alpha$ only varies between 0 and 90 deg where $\theta = 0$ deg indicates perfect alignment and $\theta = 90$ deg indicates orthogonality.

Figure \ref{orienta} shows the differences of alignment between the minor axes of the gas distribution and all semi-axis of the DM and star halos. In each panel, the difference of alignment is shown as a function of the total mass, $M_{\rm tot}$. In general, the alignment between the gas and the DM is relatively strong between minor axes, with the median alignment typically being $\alpha \simeq 40$ deg at $M_{\rm tot} = 10^{10}$ M$_\sun$ and declining to $\alpha \simeq 30$ deg at $M_{\rm tot} = 10^{11}$ M$_\sun$ down to $\alpha \simeq 20$ deg at $M_{\rm tot} > 10^{12}$ M$_\sun$. It is important to note that the differences of alignment with the less massive DM halos is imprecise since they have a quasi-spherical shape in most cases (see Sect. \ref{shapeparameters}). 

The alignment of the gas and the stars is only slightly dependent on the mass of the halos. The respective minor axes of gas and star halos have a median alignment of at least 40 deg and decrease and stabilise at around 20 deg for all masses. This behaviour will have to be confirmed with a more massive sample of halos.

\cite{2010Bet} quantified the difference of alignment angle between the stellar and total matter distribution in a sample of disc galaxies selected from a hydrodynamic simulation. They found that half of these galaxies have a difference of alignment larger than 45 deg. Using a sample of the EAGLE and cosmo-OWLS simulations, \cite{2015Velliscig} reported that the difference of alignment of massive gas halo is as large as 50 deg and dropped to 30 deg for the halo mass range $13 < \log_{10}$(M$_{200}$) $<14$. Despite the lack of data for massive halos beyond $10^{12}$ h$^{-1}$M$_\sun$ in the small EAGLE simulation, both studies are in broad agreement with our findings for similar halo masses. A second paper will follow presenting the results of larger simulations including halos of larger masses.

\subsection{Semi-axes}\label{semiaxes}
\begin{figure*}
\centering
    \includegraphics[width=5.8cm]{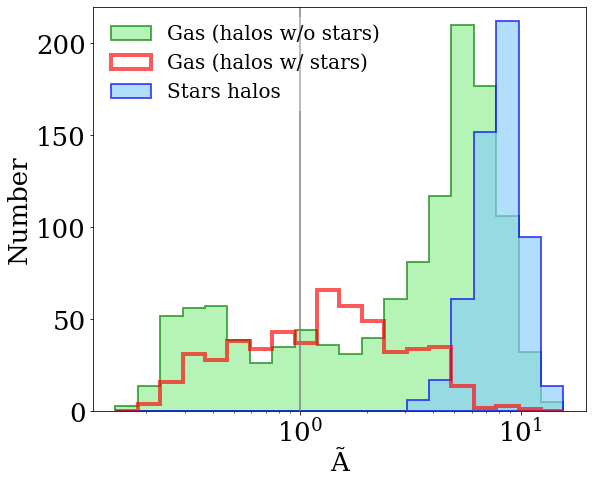}
    \includegraphics[width=5.8cm]{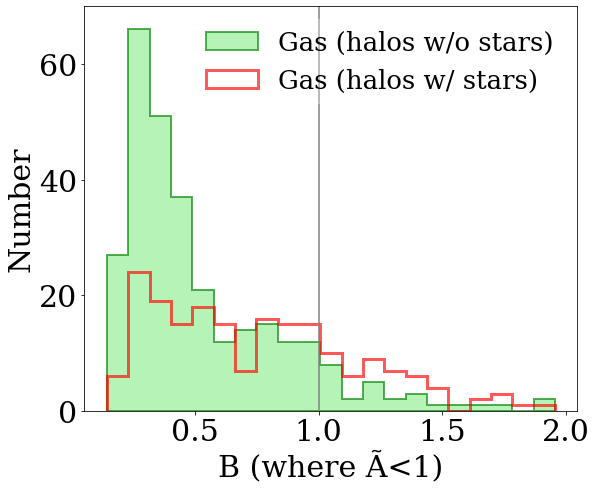}
    \includegraphics[width=5.7cm]{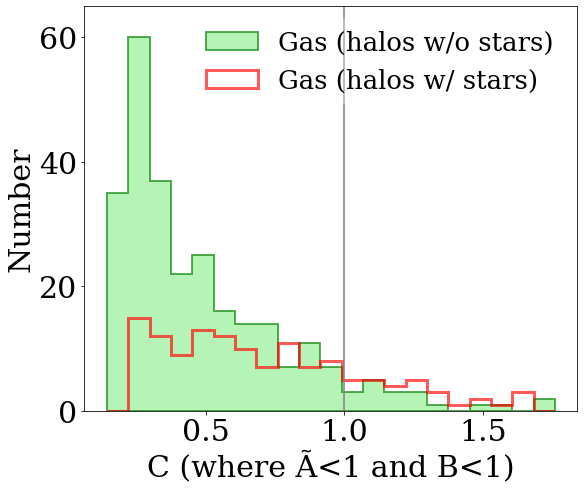}
  	\caption{Axis ratios \~{A}, \~{B} and \~{C}. Each panel corresponds to a sub-distribution. The left panel corresponds to the \~{A$_{gas}$}$=a_{DM}/a_{gas}$ (green and red bins) and \~{A$_{stars}$}$=a_{DM}/a_{stars}$ (blue bins) axis ratios. Values below 1 correspond to a semi-axis of gas (stars) larger than their DM equivalent. Middle panel corresponds to the \~{B} distribution with only the \~{A}<1 halos. Bottom panel corresponds to the \~{C} distribution with only halos with \~{A}<1 and \~{B}<1.}
  	\label{ratio_semi_dm_gas}
\end{figure*}

In Fig. \ref{ratio_semi_dm_gas}, we present the distribution of the ratio $\tilde{A}_{gas}=a_{DM}/a_{gas}$ and $\tilde{A}_{stars}=a_{DM}/a_{stars}$ of the fitted semi-axes of halos according to the type of particles. As mentioned before, gas halos are split in two groups, gas halos depleted from stars (1\,440) and gas halos with a sufficient number of stars (682). When considering all particles together, the resulting distributions are similar to those of DM, and this is why the all-particles case is not added to the figures. Since we see the same behaviour on all three axes (a,b,c), we only present the semi-axis $a$, but the results also apply for the $b$ and $c$ axes.


We remind the reader that the length of the semi-axes are obtained from the square root of the eigenvalues of the PCA. Thus, the measurement of the principal axes reflects the characteristic size in each direction of the halos. In this way, the \~{A} ratio can be used to estimate the extension of the distribution. If the semi-axis value is small, halos have small extension, whereas the larger the value, the more widespread the halo. This means that \~{A} $>1$ denotes that gas or star halos are less extended than DM halos, which is the case for star halos in Fig. \ref{ratio_semi_dm_gas}. For gas halos, we observe two cases: some of the gas halos w/o stars (26\%) and w/ stars (42\%) have larger semi-axes than the DM one. This behaviour is observed both for gas halos w/o star particles and for halos with at least 100 star particles, but it is more significant for halos with more than 100 star particles.

We then look at the ratio between the length of the gas and DM semi-axes in detail. Fig. \ref{ratio_semi_dm_gas} also presents the distribution of the ratios \~B$_{gas} = b_{dm} / b_{gas}$ (middle panel) and \~C$_{gas} = c_{dm} / c_{gas}$ (right panel) for the 2\,122 halos of gas and for each semi-axis (a,b,c). For \~{B} and \~{C}, we only consider halos with \~A$_{gas} < 1$ and \~A$_{gas}$, \~B$_{gas} <1$, respectively. We see that we have a significant number of gas halos that have major axis ratios smaller than 1. Looking at the ratios \~{B} and \~{C}, we see that the gas halos with a ratio \~{A} smaller than unity can have a ratio \~{B} greater than 1, meaning that the gas particles are less extended than their DM counterpart along this axis. However, this does not represent the majority of halos. A significant proportion of halos (71\%) that have a ratio on \~{A} < 1 also have a ratio on \~{B} < 1 and \~{C} < 1.

\subsection{Shape of halos}\label{shapeparameters}
\begin{figure}
    \resizebox{\hsize}{!}{\includegraphics{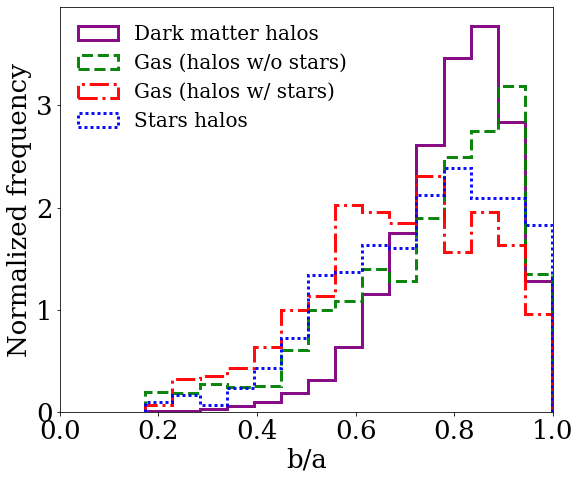}}
    \resizebox{\hsize}{!}{\includegraphics{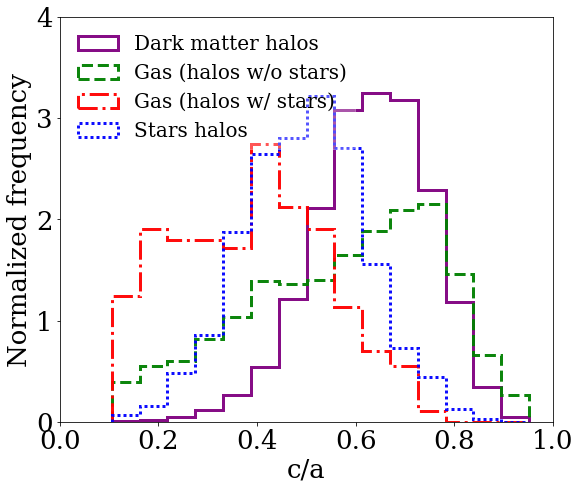}}
  	\caption{Distribution of $b/a$ (top panel) and $c/a$ (bottom panel) axis ratios according to the type of particle contained in the RefL0025N0376 simulation at z=0.5. DM halos (17\,480) are represented in pink. Gas halos are split in two groups: gas halos w/o stars (1\,440) in green and gas halos w/ stars (682) in red. Halos of stars (682) are represented in blue.}
    \label{axis_ratio_types}
\end{figure}

In Fig. \ref{axis_ratio_types}, we present the distribution of the axis ratios $b/a$ and $c/a$ of the 17\,480 halos from the RefL0025N0376 simulation according to the types of particles they contain (DM, gas, or stars). Table \ref{axis_ratio_table} summarises the median characteristics of the distributions of axis ratios according to the particle types.

\begin{table}[!htp]
\centering
\caption{Analysis of the 17\,480 halos in the RefL0025N0376 simulation at z=0.5. Median and median-absolute deviations of axis ratios $b/a$ and $c/a$ are provided according to the type of particle contained in the halos: DM (17\,480 halos), gas (1\,440 halos w/o stars, 682 w/ stars), and stars (682 halos).}
\label{axis_ratio_table}
\begin{tabular}{lllll}
\hline
  &  DM &  Gas  & Gas &  Stars \\
  & & (w/ stars) & (w/o stars) & \\
\hline
b/a  &  $0.82 \pm 0.11$ & $0.70 \pm 0.19$ & $0.80 \pm 0.16$ & $0.76 \pm 0.19$ \\
c/a  &  $0.64 \pm 0.12$ & $0.38 \pm 0.20$ & $0.60 \pm 0.22$ & $0.50 \pm 0.12$\\
\hline
\end{tabular}
\end{table}

The EAGLE halos at z=0.5 are described by a median axis ratio: $b/a\sim0.8$. The sphericity peaks around $c/a\sim0.6$, but we notice that the gas component presents an over-representation at small values of sphericity $c/a$ corresponding to flatter ellipsoids of gas. These flatter halos coincidentally correspond to halos with $N_{stars}>100$. DM halos have weak ellipticity with strong sphericity, as $b/a \sim 0.8$ and $c/a \sim 0.7$. Because the EAGLE simulation's prescription for star formation uses a Jeans-mass-limiting pressure floor, it is difficult for gas to cool into thin discs before creating stars, resulting in an artificial thickening of the stellar component and perhaps explaining why there are very few star halos with $c/a < 0.3$. Despite the fact that we used a different approach based on a PCA, the median values of the DM, gas, and star sphericity are largely consistent with those retrieved by \cite{2015Velliscig,2016Shao}, and \cite{hill2021} for halos in the same mass range. Similarly, the distribution of $b/a$ and $c/a$ axis ratios are consistent with those recovered by \cite{2014Tenneti} in the MassiveBlack-II and by \cite{2019Illustris50} in the IllustrisTNG simulations.

\begin{figure*}
\centering
    \includegraphics[width=5.8cm]{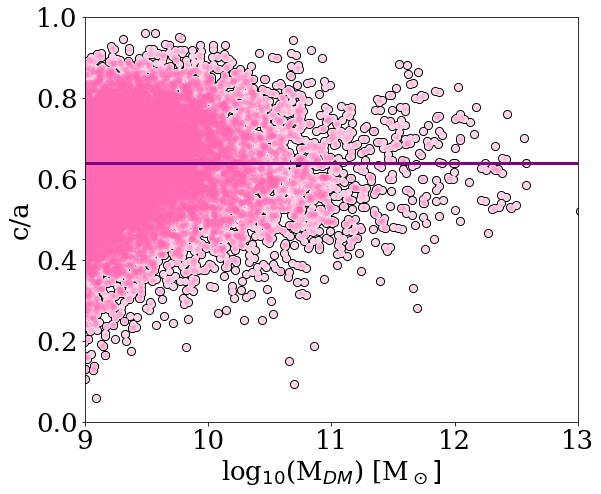}
    \includegraphics[width=5.8cm]{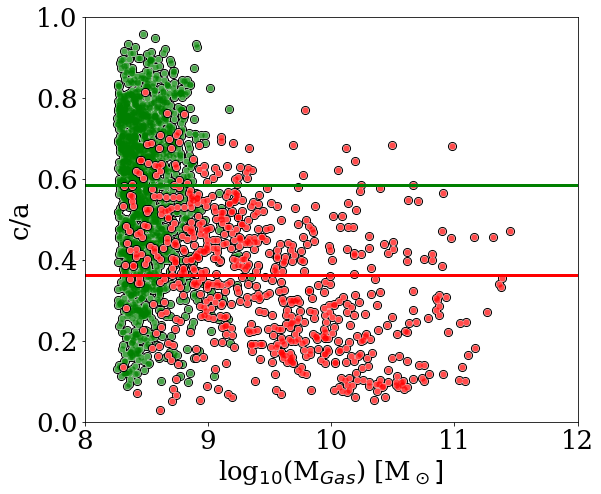}
    \includegraphics[width=5.8cm]{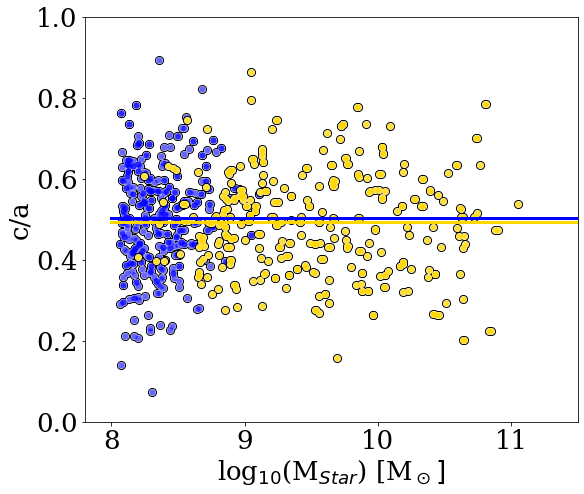}
  	\caption{Analysis of 17\,480 halos at z=0.5 of RefL0025N0376 EAGLE simulation. The distribution of the sphericity, $c/a$, is shown as a function of the total mass of the halos for DM (left panel), gas (middle panel), and stars (right panel). Gas halos are split in two groups: gas halos w/o stars (1\,440) are in green and gas halos w/ stars (682) are in red. Star halos are split into two groups of approximately equal size: low-mass halos ($M_{tot} < 10^{11}$ M$_{\sun}$), depicted by blue dots, and high-mass halos ($M_{tot} > 10^{11}$ M$_{\sun}$), depicted by yellow dots. Colour lines represent the median.}
    \label{sphericity_mass_types}
\end{figure*}


We now investigate the dependence of the sphericity on the mass of the halos, with a particular focus on gas halos. Fig. \ref{sphericity_mass_types} presents the distribution of the sphericity, $c/a$, for DM, gas (split in w/ stars and w/o stars), and star halos. We split star halos into two groups: low-mass halos ($ M_{tot} < 10^{11} M_\sun$) and high-mass halos ($ M_{tot} > 10^{11} M_\sun$). This threshold was chosen so as to have 50\% of the star halos in each group.

There is a clear dependency on the sphericity of the gas halos with the mass of the halo: the less massive halos tend to have a more spherical distribution of gas with median value of $c/a \sim 0.6$, whereas for high-mass halos ($M_{\rm tot}>10^{11}\,$M$_\sun$), the sphericity has a median value of $c/a \sim 0.4$. Star halos do not follow the same behaviour since we do not observe any influence of mass on the flattening. The lack of massive star halos does not allow us to draw a definitive conclusion. The DM sphericity is totally insensitive to the mass of the halos.

\subsection{Mass distribution of halos}\label{massdistributionsec}
\begin{figure}
    \resizebox{\hsize}{!}{\includegraphics{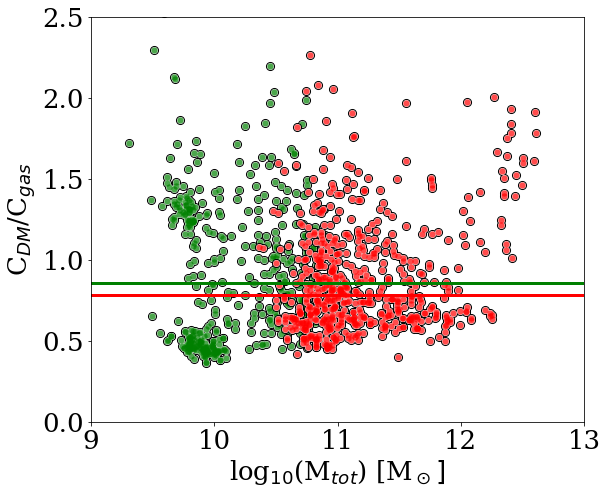}}
    \resizebox{\hsize}{!}{\includegraphics{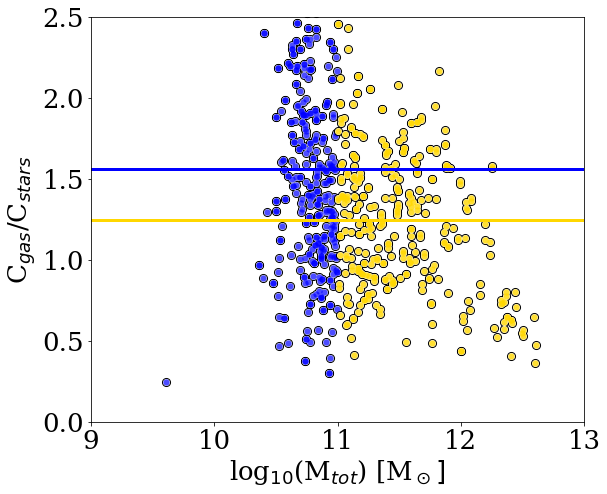}}
    \caption{DM-to-gas (top panel) and gas-to-star halo (bottom panel) concentration ratios as a function of their total mass for 1\,030 halos in the RefL0025N0376 simulation. Halos are composed of a minimum of 200 particles. Gas halos are split in two groups: gas halos w/o stars (419) in green and gas halos w/ stars (611) in red. Star halos are split into two groups of approximately equal size: low-mass halos ($M_{tot} < 10^{11}$ M$_{\sun}$), depicted by blue dots, and high-mass halos ($M_{tot} > 10^{11}$ M$_{\sun}$), depicted by yellow dots. Colour lines represent the median.}
    \label{concentr}
\end{figure}

In addition to the shape and orientation parameters, it is also interesting to investigate the mass distribution in the halos, which gives insight into their internal structure. For this we used the concentration parameter $C$ from Equation \ref{Concentration}, which relies on the mass growth curve. Fig. \ref{concentr} presents the ratio of the concentration parameters between DM, gas, and star halos as a function of their total mass. In order to avoid poorly constrained concentration estimates, we remove all halos with fewer than 200 particles.

We observe in the top panel of Figure \ref{concentr} a concentration ratio $C_{DM}/C_{gas}$ slightly smaller than 1 for halos with and without a sufficient number of stars. This shows that the gas of these halos is slightly more concentrated than the DM.

\begin{figure}
    \resizebox{\hsize}{!}{\includegraphics{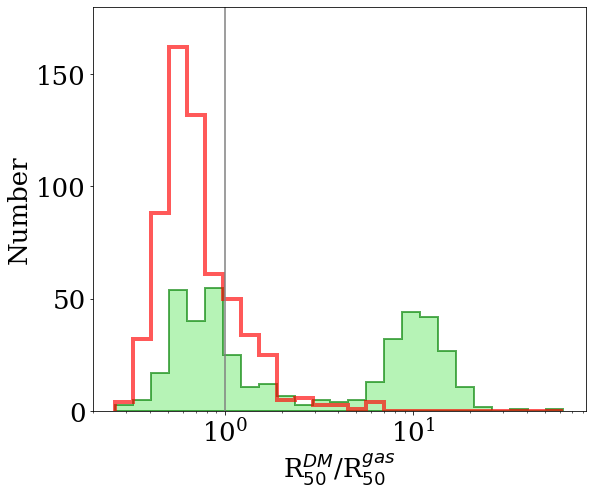}}
    \resizebox{\hsize}{!}{\includegraphics{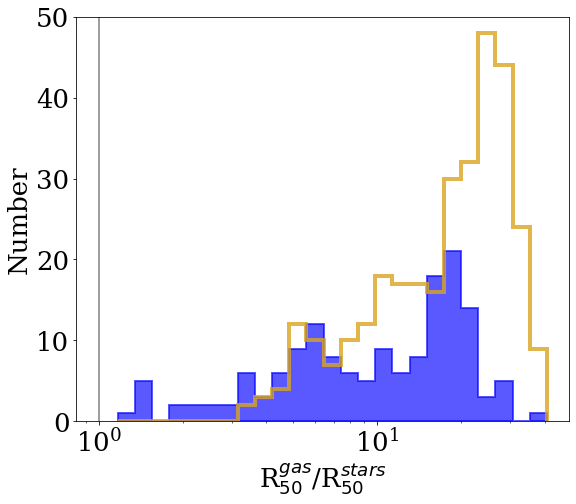}}
  	\caption{Distribution of the ratio between the half-mass radius of DM and gas halos (top panel) and gas and star halos (bottom panel). Halos are composed of a minimum number of 200 particles. Colour scheme is similar to that of Figures \ref{sphericity_mass_types} and \ref{concentr}. The constant lines R$^{DM}_{50}$/R$^{gas}_{50}$=1 and R$^{gas}_{50}$/R$^{stars}_{50}$=1 are also plotted as references.}
  	\label{hmr_ratio}
\end{figure}

Regarding the concentration of star particles compared to gas particles, we observe that most of the star halos (69\%) are less concentrated than gas halos at redshift z=0.5. Looking at the median of the concentration ratios for the two different mass ranges, we see that the more massive stellar halos (in yellow) are more concentrated than the less massive ones (in blue). It is important to remember here that the concentration does not provide a measure of the extent of the halos. Therefore, although the gas may be more concentrated than the stars, the latter are effectively formed in the core of the gas halos as shown in Fig \ref{hmr_ratio}. In this figure, we show the distribution of the ratio between the half-mass radius of DM and gas halos (top panel) and gas and star halos (bottom panel), respectively.

The distribution of the ratio of this effective radii according to the type of particles shows that the star particles are well concentrated in the core of the halos. There is no stellar halo that has an effective half-mass radius larger than its corresponding gas halo. The EAGLE simulation therefore contains gas halos that are more diffuse than their DM counterparts.

\subsection{Properties of the projected halos}\label{properties_projected_halos}
\begin{figure}
    \resizebox{\hsize}{!}{\includegraphics{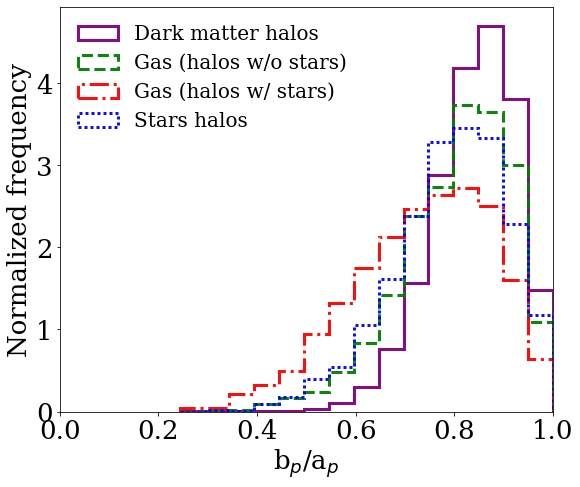}}
  	\caption{Unit-area distribution of projected axis ratios of 61 398 projections from the RefL0025N0376 simulation at z=0.5 according to the particle type: DM (61\,398 halos), gas (4\,362 halos w/o stars, 2\,058 w/ stars), stars (2\,172 halos).}
  	\label{projected_ba}
\end{figure}

The work presented in previous sections investigates the three dimensional shape and mass distribution of halos from the RefL0025N0376 (z=0.5) simulation. Instead, common analyses of galaxy morphology using imaging surveys usually derive parameters characterising the properties of projected halos onto the sky plane. Here, we analyse the properties of the EAGLE halos when projected onto the plane of the sky, allowing a comparison with ground-based and space-based imaging surveys and giving a measurement of the gravitational lensing effect that gives, in turn, constraints on the matter distribution along the line of sight. For commodity and under the reasonable assumption that halos are uniformly oriented, we use the coordinates of the particles in the XY, YZ, and XZ planes as the projections of the halos in three directions. Fig. \ref{projected_ba} presents the axis ratio ($b_p/a_p$) of the projected halos from the RefL0025N0376 (z=0.5) EAGLE simulation. Table \ref{table_projected} summarises the median characteristics of the projected distributions of axis ratios according to the particle type. 

\begin{table}[!htp]
\centering
\caption{Median and median-absolute deviations of the projected axes ratios $b_p/a_p$, according to the type of particle: DM (61\,398 halos), gas (4\,362 halos w/o stars, 2\,058 w/ stars), and stars (2\,172 halos).}
\label{table_projected}
\small{
\begin{tabular}{lllll}
\hline
  &  DM &  Gas  & Gas &  Stars \\
  & & (w/ stars) & (w/o stars) & \\
\hline
$b_p/a_p$  &  $0.85 \pm 0.06$ & $0.75 \pm 0.10$ & $0.82 \pm 0.08$ & $0.80 \pm 0.07$ \\
\hline
\end{tabular}
}
\end{table}

The overall distribution of the axis ratios from DM has a median value of $(b_p/a_p)_{DM} = 0.85\pm0.06$ corresponding to quasi-spherical halos. Gas and stars have similar properties with a little more elongated projected profiles with axis ratio values of $(b_p/a_p)_{gas} = 0.75\pm0.10$ for gas halos with more than 100 star particles and $(b_p/a_p)_{gas} = 0.82\pm0.08$ for gas halos with fewer than 100 star particles. Star halos have a median axis ratio of $(b_p/a_p)_{stars} = 0.80\pm0.07$. The three normalised distributions agree at a $1\sigma$ level.

The median value of the axis ratios for stars can be compared to the morphology derived in imaging surveys such as the SDSS DR16 \citep{2020Ahumada} or derived from the space-based Gaia observations \citep{2022Ducourant, 2022Bailer-Jones}. We analysed the distribution of the axis ratios of 390\,612 selected galaxies from the SDSS DR16 for which the parameters of a de Vaucouleurs profile are provided. The mean value of their axis ratios is $(b/a)_{SDSS}=0.65\pm0.21$. The SDSS estimates describe stellar content of galaxies with more elongated shape than what is observed in the EAGLE simulation at z=0.5. The same sample of galaxies, but observed by Gaia with a de Vaucouleurs profile, has the following axis ratio values: $(b/a)_{Gaia}\sim0.73\pm0.14$. Depending on the survey ground-based or space-based the mean values of an axis ratio are somewhat different, but they are in good agreement one with the other and with the EAGLE value, within 1$\sigma$.

\begin{figure}
    \resizebox{\hsize}{!}{\includegraphics{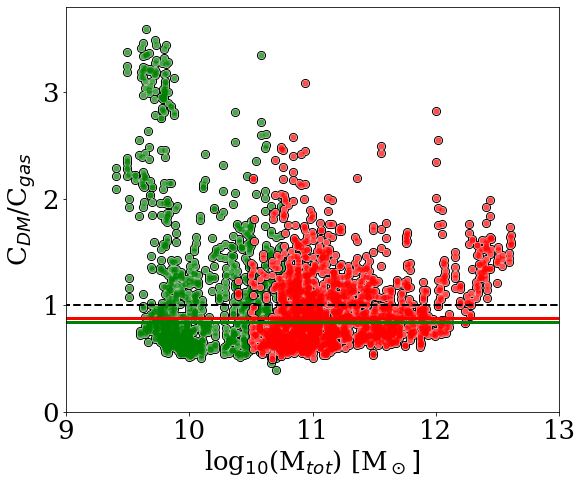}}
    \resizebox{\hsize}{!}{\includegraphics{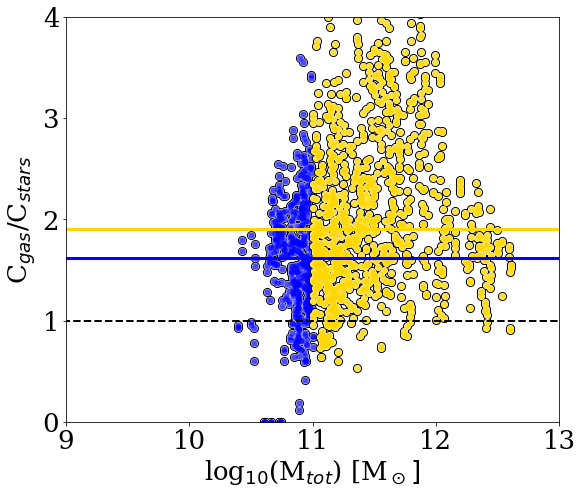}}
  	 \caption{Concentration ratio of the projected halos at z=0.5 from the RefL0025N0376 simulation. The halos have a minimum of 200 particles. The ratios $C_{DM}/C_{gas}$ (top panel) and $C_{gas}/C_{stars}$ (bottom panel) are expressed as a function of the total mass of the halo. Gas halos are split in two groups, gas halos w/o stars (1290) in green and gas halos w/ stars (1875) in red. Star halos are split into two groups of approximately equal size: low mass halos ($M_{tot} < 10^{11}$ M$_{\sun}$), depicted by blue dots, and high mass halos ($M_{tot} > 10^{11}$ M$_{\sun}$), depicted by yellow dots. Colour lines represent the median. The constant dashed lines C$_{DM}$/C$_{gas}$=1 and R$_{gas}$/R$_{stars}$=1 are also plotted as references.}
  	\label{projected_concentration_mass}
\end{figure}

Figure \ref{projected_concentration_mass} presents the ratio of concentration of the projected DM and gas halos (top panel) and gas and star halos (bottom panel) as a function of the total mass.

In it, we see a $C_{DM}/C_{gas}$ concentration ratio slightly smaller than 1 between DM and gas. This shows that the gas is slightly more concentrated than the DM. This result is consistent with the observations from Section \ref{massdistributionsec}. Star halos have a lower concentration than gas. This behaviour is the opposite of what is observed for the three dimensional halos. 

\section{Dependence on the redshift}\label{dependredshift}
The RefL0025N0376 simulation contains 29 snapshots with different redshifts varying from $z=15$ down to $z=0$. The analysis of the number of halos, as well as their morphology and mass distribution, at the different redshifts allows an insight into the cosmic evolution. In the rest of this study, we chose to represent the evolution of the shape of the halos as a function of the look-back (LB) time. This change of scale between redshift and LB time allows us to have a linear time scale in the EAGLE snapshots and to focus more on the near Universe. We used the same cosmological parameters as the EAGLE simulation for the scale conversion, as provided in Section 2.

\subsection{Frequencies}\label{freq}
\begin{figure*}
\centering
    \includegraphics[width=5.8cm]{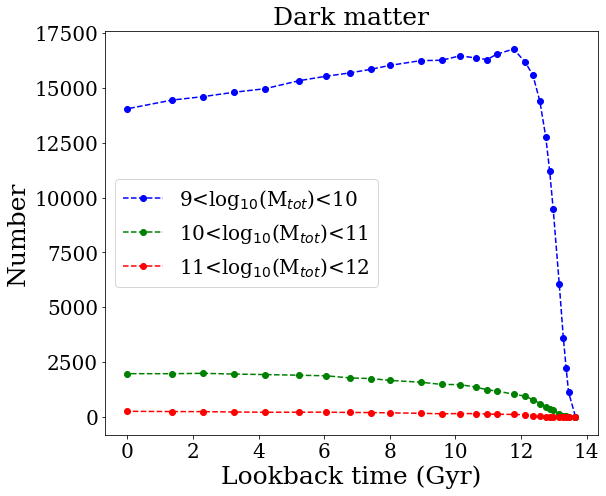} 
    \includegraphics[width=5.8cm]{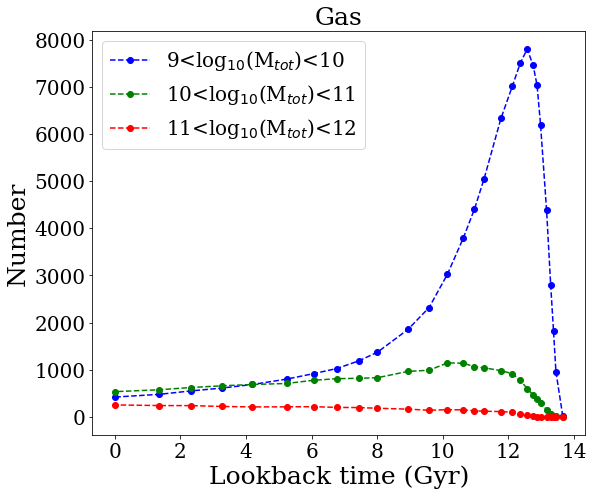}
    \includegraphics[width=5.8cm]{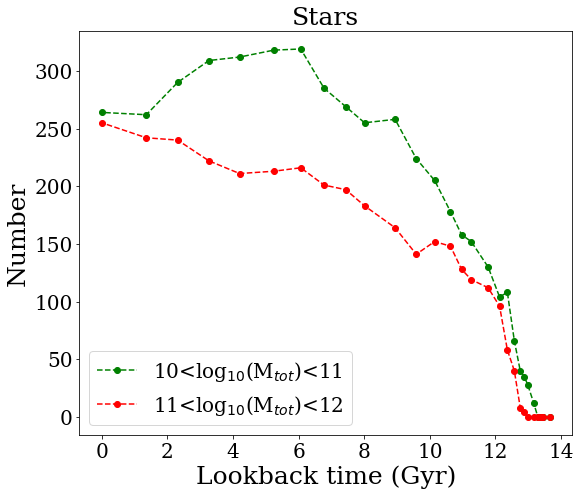}
  	\caption{Variation of number of halos with at least 100 particles from the RefL0025N0376 simulation as a function of redshift. Halos are grouped according to the type of their constituent particles and split in various ranges of solar masses.}
    \label{redshift_number_halos}
\end{figure*}

The number of halos is not constant with redshift and the presence of stars is linked with the stage of evolution of the halos. In a $\Lambda$-CDM universe, high-mass DM halos are formed more recently than their low-mass counterparts \citep{2017Yu}. This is confirmed in Fig. \ref{redshift_number_halos}, which presents the number of halos at various redshifts and according to the particle type for the RefL0025N0376 simulation. The number of low-mass DM halos increases with time. The number peaks at $LB = 12.13$ Gyr (z=3.53) and then decreases. The rise illustrates that the simulation becomes more and more contrasted (the FoF algorithm detects more and more halos) as structures emerge. From $LB=10.15$ Gyr (z=1.74), we mainly observe halo mergers in hierarchical mass growth.

We observe a similar behaviour for gas halos at $LB = 12.76$ Gyr (z=5.04), where the number of low-mass halos peaks and then decreases. Low-mass gas halos that formed at high redshifts do not exist in the present day and merged into more massive halos \citep{2017Yu}.

\subsection{Morphology}\label{morphi}

\begin{figure}
    \resizebox{\hsize}{!}{\includegraphics{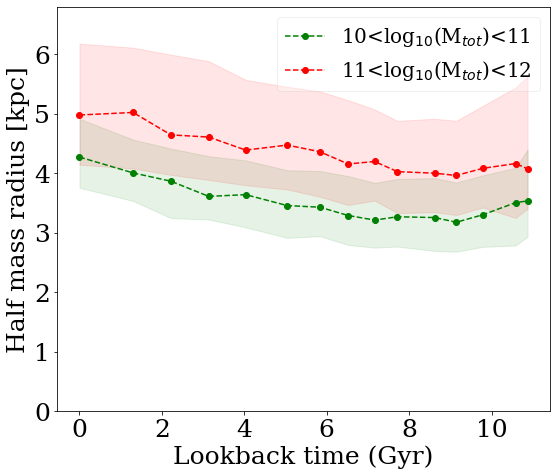}}
  	\caption{Median of the half mass major-axis radius of the star halos as a function of the LB time for various solar mass ranges. The solid lines indicate the median values, whilst the shading denotes the 25th-75th quartiles.}
    \label{size_redshift}
\end{figure}

Figure \ref{size_redshift} presents the evolution of the half mass radii of the star halos from the RefL0025N0376 simulation as a function of the LB time for two mass ranges. 

One first observes that halos more massive than $10^{10}$M$_\sun$ appear from $LB = 13.17$ Gyr ($z=7.05$) and more massive than $10^{11}$M$_\sun$ from $LB = 12.88$ Gyr ($z=5.49$) corresponding to the respective time scales necessary for significant fusions of halos producing sufficient stars. The size of the star halos grows with time whatever their mass range is. The growth rate is similar. The growth of halo size results from the merging along time of halos to produce more massive and hence larger halos and by the formation of new stars. 

\begin{figure*}
\centering
    \includegraphics[width=5.8cm]{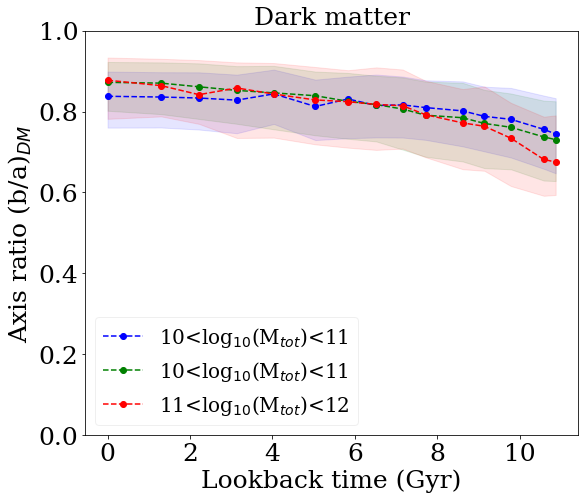} 
    \includegraphics[width=5.8cm]{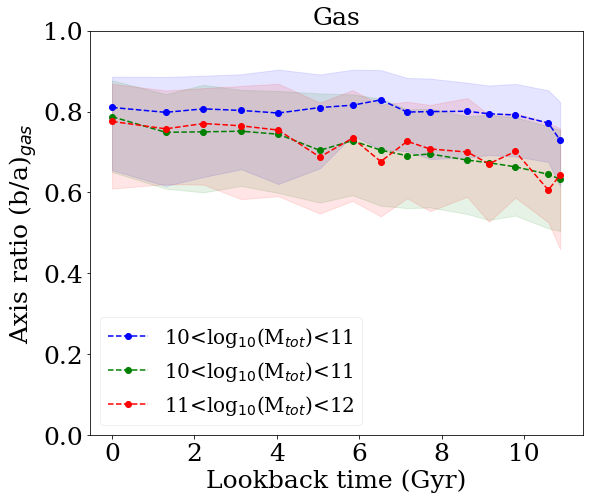}
    \includegraphics[width=5.8cm]{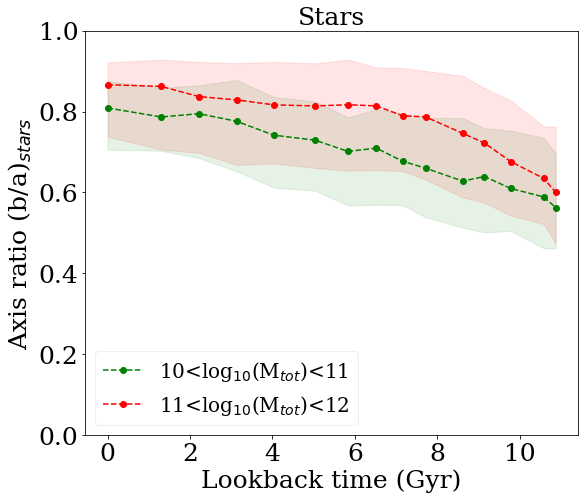}
  	\caption{Dependence of the median of axis ratios ($b/a$) on the LB time for halos in the RefL0025N0376 simulation according to various solar mass ranges and particle types.}
    \label{redshift_axis_ratio_ba}
\end{figure*}

\begin{figure*}
\centering
    \includegraphics[width=5.8cm]{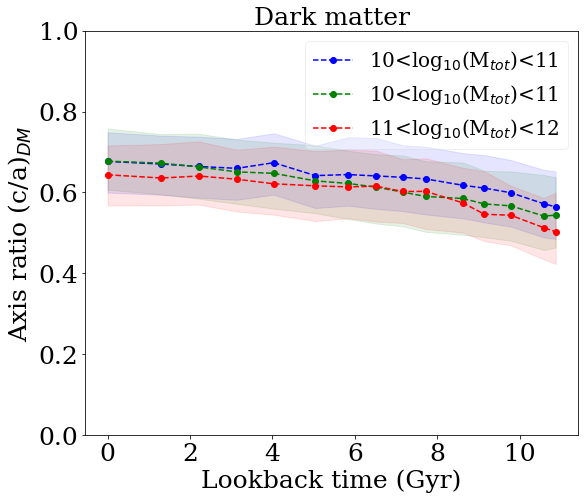} 
    \includegraphics[width=5.8cm]{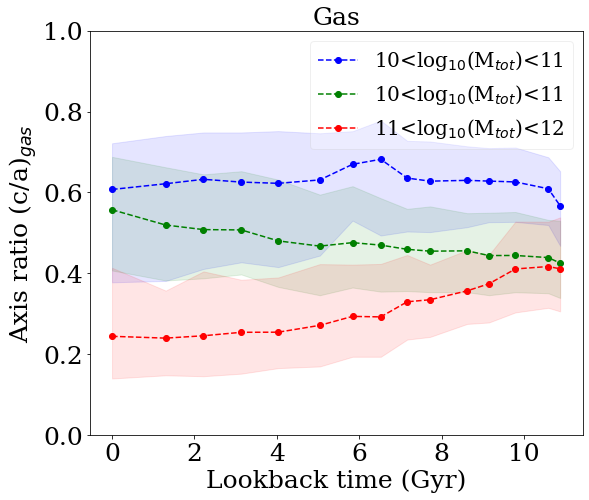}
    \includegraphics[width=5.8cm]{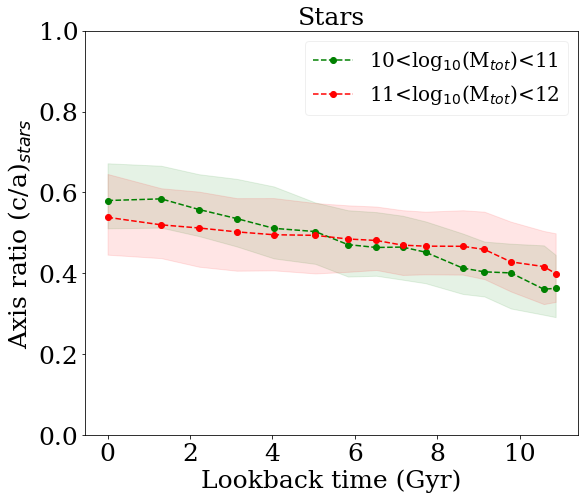}
  	\caption{Dependence of the median of axis ratios ($c/a$) (sphericity) on the LB time for halos in the RefL0025N0376 simulation according to various solar mass ranges and particle types.}
    \label{redshift_axis_ratio_ca}
\end{figure*}

In order to investigate the mass dependence on the distribution of the axis ratio, we plot in Figure \ref{redshift_axis_ratio_ba} and \ref{redshift_axis_ratio_ca}, the median of the axis ratios ($b/a$) and ($c/a$) for the mass bins $10^{9} \leq M_{\rm tot}<10^{10}$ M$_\sun$, $10^{10}\leq M_{\rm tot} <10^{11}$ M$_\sun$, and $10^{11}\leq M_{\rm tot} < 10^{12}$ M$_\sun$.


The shape of the halos when all particles are considered is dominated by the DM component if all types of particles are considered. In fact, the shape of the DM component is nearly identical to the total mass distribution, so we did not consider the case of all particles. The global tendency of the axis ratios is to become more and more spherical with time, whatever the mass and the particle type are. These plots can be compared against \cite{2006Allgood} and \cite{2014Tenneti}. Our results agree with theirs for DM halos in the sense that the average axis ratios, $b/a$ and $c/a$, increase as we get closer to the present time ($LB$=0). This effect is not due to baryon physics since it is also found in DM-only simulations \citep{2013Bryan,2014Tenneti}. The halos are built up by accreting matter from the cosmic web filaments, resulting in an initial anisotropy, with a progressive virialisation that leads to a final ellipsoid for these non-colliding particles that approach a spherical halo.

For the stellar matter, we see that the average axis ratios generally follow the DM ratios trends. Star particles are, as DM particles, non-colliding. They behave as massless particles in the gravitational potential defined by the DM-dominated halo. The only exception concerns the very massive gas halos (M$_\sun>10^{11}$), for which an inverse tendency for ($c/a$) is observed. These halos' discs tend to become more and more flattened with time. For gas, low-mass halos (M$<10^{10}\,$M$_\sun$) have larger axis ratios than more massive halos. This is correlated with the absence of stars in the low-mass halos.

Compared to the previous b/a ratio, the c/a axis ratio is slightly smaller than b/a. This results in rather flattened halos. The c/a axis ratio of the gas shows a strong dependence on the mass. The greater the mass, the more the axis ratio tends towards small values of c/a. This shows that gas halos tend to flatten out as they gain mass. The minimum mass observed for a halo of stars to form ($N_{stars}>100$) is $M \sim 10^{11}$ M$_\sun$, revealing that a necessary mass level is required to form enough stars.

\subsection{Concentration as function of redshift}
\begin{figure}
    \resizebox{\hsize}{!}{\includegraphics{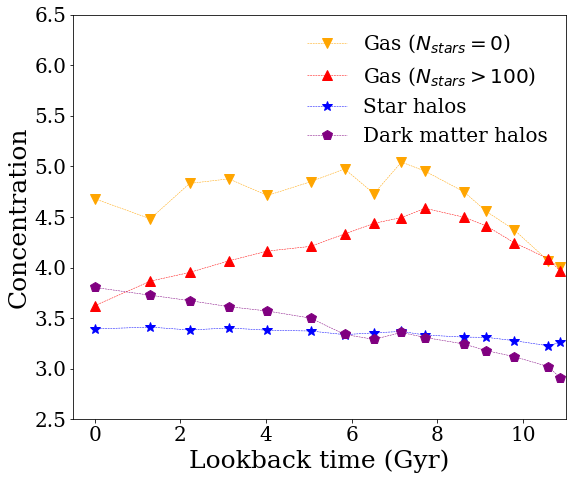}}
  	\caption{Concentration as a function of LB time for RefL0025N0376 simulation. The solid lines indicate the median values. Gas halos are split in two groups, gas halos without any stars in yellow and gas halos with more than 100 star particles in red.}
    \label{concentration_gas_redshift}
\end{figure}

As in Section \ref{properties_projected_halos}, we remind the reader that we did not consider halos with fewer than 200 particles for concentration measurements. Fig. \ref{concentration_gas_redshift} presents the evolution of the median of the concentration as a function of the LB time.

The concentration of DM and stars increases continuously during the simulation. The DM concentration increases more strongly at the beginning of the simulation ($10<T_{lookback}<13$ Gyr) than at more recent times ($0<T_{lookback}<8$ Gyr). The DM concentration increases as halos form and as these new halos accumulate mass by accreting particles into their gravitational pool. Then, the gas particles follow the gravitational potential and fall in the core of the DM halos. This is observed through a significant increase of the gas concentration in the first gigayear of the simulation up to an LB time of $\sim 8$ Gyr, where the concentration peaks at around $4.5$ for halos that have already formed some stars. This reflects the fact that high-concentration gas halos slowly lose their concentration over time. This can have two possible origins. The first is that the concentrated gas in the heart of the halos is consumed to fuel the formation of new stars. Then, the gas is dispersed by these new stars formed in the heart of the halos. Gas halos that do not form any star particles are small halos ($M_{\rm tot}<10^9$ M$_\sun$) and have a very low decreasing concentration as no star consumes the gas.

\subsection{Projected axis ratio}

\begin{figure*}
\centering
    \includegraphics[width=5.8cm]{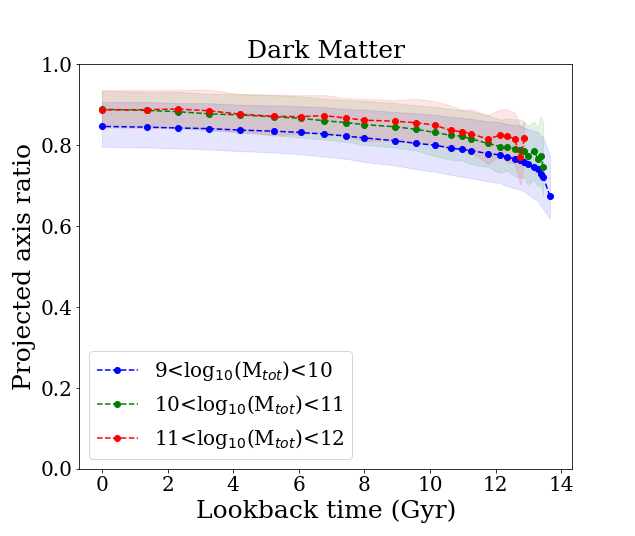} 
    \includegraphics[width=5.8cm]{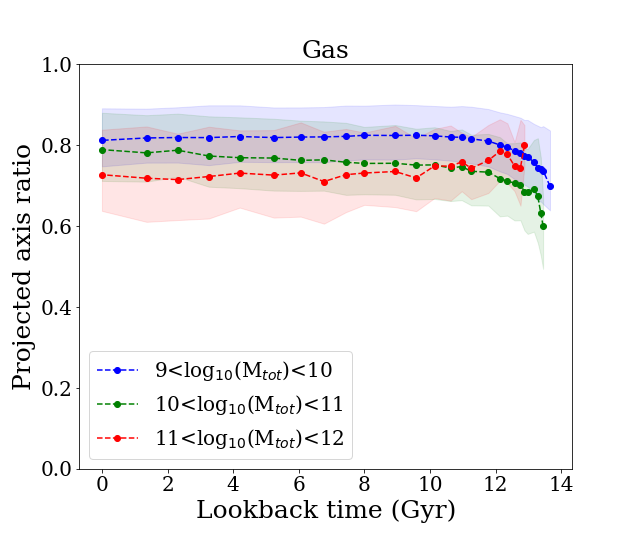}
    \includegraphics[width=5.8cm]{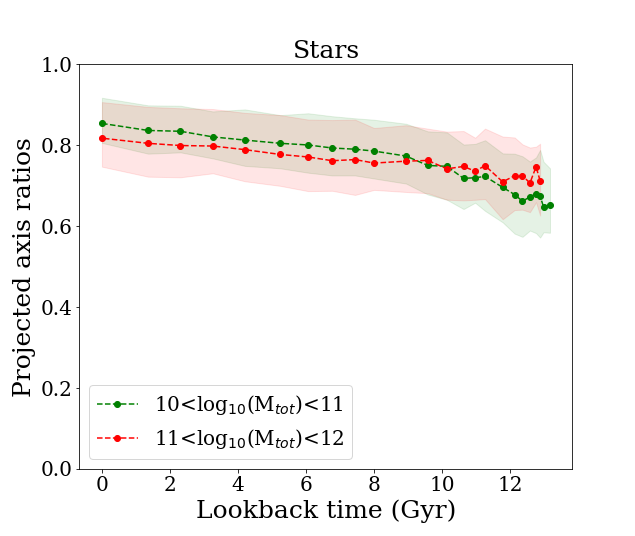}
  	\caption{Dependence of median of projected axis ratios ($b_p/a_p$) on the LB time for halos in RefL0025N0376 simulation according to various solar mass ranges and particle types.}
    \label{redshift_axis_ratio_projected}
\end{figure*}

In this section, we repeat the analysis performed for the three dimensional halos in Sections \ref{freq} and \ref{morphi} for the two dimensional projected halos. Fig. \ref{redshift_axis_ratio_projected} presents the projected axis ratio as function of the redshift for each type of particle and by mass range.

We observe that the halos are more spherical on average for low values of the LB time. DM has a constantly increasing distribution. The greater the mass, the greater the axis ratio of the halo. The general evolution of stars is identical to that of DM. For star particles, there is an inversion in the evolution of the most massive halos. For the most massive halos, $M_{\rm tot}>10^{11}$ M$_\sun$, the axis ratio of the stars tends to be lower. For gas halos of $M>10^{11}\,$M$_\sun$, the projected axis ratio has a slight tendency to decrease.

\subsection{Projected concentration}

\begin{figure}
    \resizebox{\hsize}{!}{\includegraphics{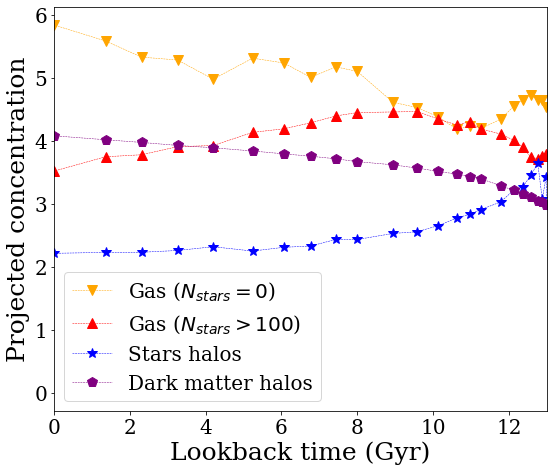}}
  	\caption{Projected concentration as a function of the LB time for the RefL0025N0376 simulation. The lines indicate the median values. Gas halos are split in two groups, gas halos without any stars in yellow and gas halos with more than 100 star particles in red.}
    \label{concentration_gas_redshift_mass}
\end{figure}

Figure \ref{concentration_gas_redshift_mass}  presents the evolution of the median of the projected concentrations as a function of the LB time. The behaviour of the DM particles is identical to that already observed on the concentration measured along the main axes. The concentration of the DM halos increases continuously from three to four. The gas particles have two different behaviours that follow the previously observed trends. The gas halos containing stars see their projected concentration increase at the beginning of the simulation until they reach a peak around 8-9 Gyr. From this time, the projected concentration of the gas particles decreases continuously from a maximum of 4.5 to 3.2 at present time. This behaviour indicates the presence of star formation in the gas halos. Stars form in the centre of the halos by consuming some of the gas and then diffusing it as they go. At the beginning of the simulation, the stars, which form mainly in the heart of halos where the gas concentration is sufficient, have a concentration close to that of the gas from which they originate. Then, as more stars form, the projected concentration decreases as these particles diffuse into the halos. 

\section{Conclusion}\label{Ccl}

The EAGLE project represents a judicious test bed to obtain statistics on the mass distribution of halos, as the suite of simulations were calibrated to reproduce the galaxy stellar mass function and the size-mass relation of late-type galaxies. In this study, we used the small volume RefL0025N0376 simulation to directly measure the shapes and alignments of DM, gas, and star halos. One motivation is to obtain precise knowledge of the projected mass distribution of the deflectors in strong lensing. The shape parameters have been calculated for the 28 available redshift snapshots of the simulation with a special focus at $z=0.5$. The method we developed analyses the three dimensional particle distribution within an iteratively determined radius excluding large-scale structures, which usually perturb the processing. This radius is set so that $\sim 80\%$ of the mass of the halos is considered. 

At $z=0.5$, the minor axis of the gas halos preferentially aligns with the minor axis of the DM halos. The mean difference in alignment between the minor axes of the gas and of the DM distributions varies from $\sim 40$ deg to $\sim 20$ deg for masses of $10^{10}<M_{\rm tot}<10^{12}$ M$_\sun$. The differences in alignment between the gas and stellar components are larger than those between DM and gas halos. These results are in good agreement with previous studies \citep{2015Velliscig,thob2019,hill2021}. Since the number of massive halos is small in the simulation we studied, further analyses on the larger EAGLE simulations have already been started.

Concerning the axis ratio of the halos, the $b/a$ for all particles have a median value of $0.81 \pm 0.07$, and the sphericity $c/a$ peaks at a median value of $0.64 \pm 0.08$. The distribution of the axis ratios of the various components behave similarly, except for gas halos that already form stars and are much more flattened. These halos are more massive than the gas halos that did not form stars. 

We also measured the distribution of the projected axis ratio $b_{p}/a_{p}$ when the halos are projected on 2D sky maps. Whatever the component under consideration is, the distribution of $b_{p}/a_{p}$ has a median value of $0.86 \pm 0.05$ at $z=0.5$. As observed for the three dimensional case, this axis ratio has smaller $b_{p}/a_{p}$ values for gas halos w/ stars. 

We find that gas halos that form stars have larger concentration values than gas halos not forming stars. When comparing star halos to their gas halo counterparts, we notice that most of the star halos (69\%) are less concentrated and that this concentration depends on the total mass of the halos.

The value of the concentration parameter of DM and star halos increases continuously from redshift $z=15$ to $z=0$. Gas halos that do not form any star appear to be light halos ($M_{\rm tot}<10^9$ M$_\sun$). Their concentration increases continuously with time as no star consumes the gas. We find that the total matter distribution in halos is more spherical at higher halo masses and redshifts. The same qualitative trends hold for the star and gas distributions of these halos.

The halos show different rates of variation for each shape parameter depending on the component considered. Our analysis confirms recent observations on the shape and formation of halos. Each component evolves in a particular way with respect to the others, depending on whether the matter is collisional or not. The evolution of gas and star particles are related since stars are derived from the gas. These components are evolving rapidly, forming a diversity of structures, as observed in the current large surveys. On the other hand, DM halos evolve slowly and tend to retain traces of the initial conditions, resulting in more spherical shapes.

\begin{acknowledgements}
French CNRS INSU
PNGram
SNOGaia
CS-OASU

\cds{This project has received funding from the European Research Council (ERC) under the European Union’s Horizon 2020 research and innovation programme (grant agreement No 787886).}

\end{acknowledgements}

\bibliographystyle{aa}
\bibliography{bibliography}

\begin{appendix}
\section{Influence of the number of particles on shape characterisation}\label{appendixvalidation}

\begin{figure*}
\centering
    \includegraphics[width=5.8cm]{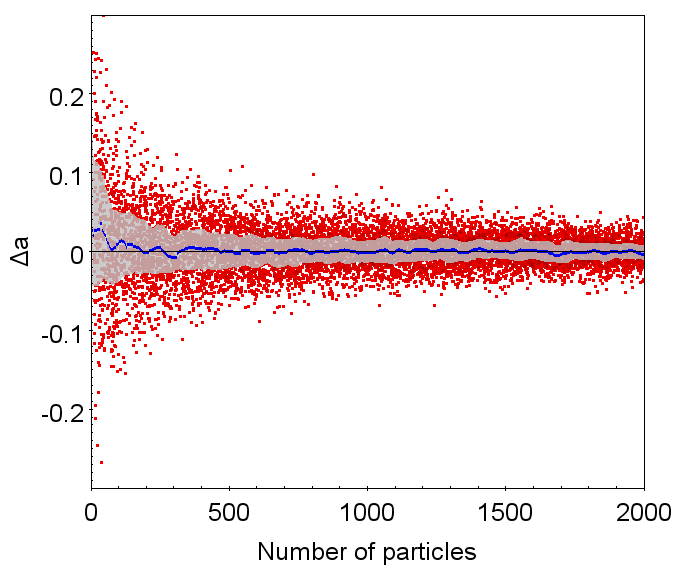}
    \includegraphics[width=5.8cm]{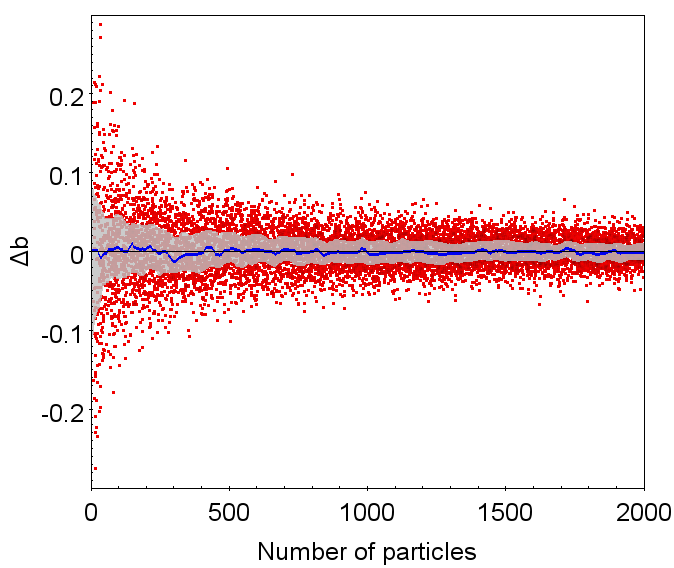}
    \includegraphics[width=5.8cm]{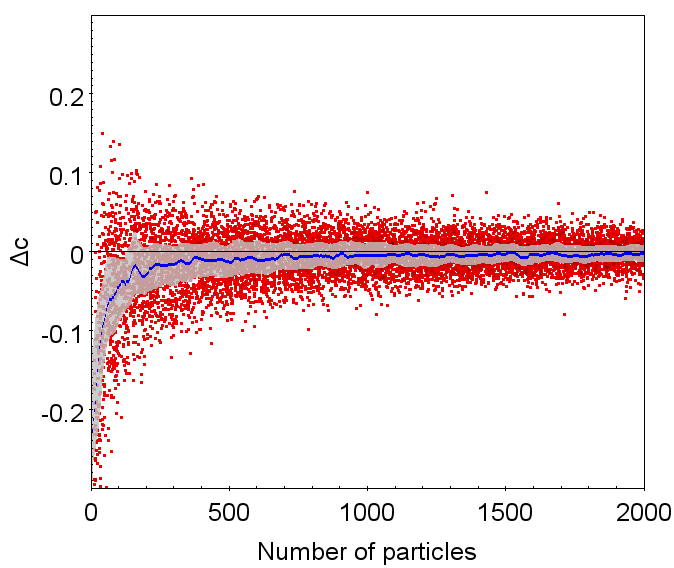}
    \includegraphics[width=5.8cm]{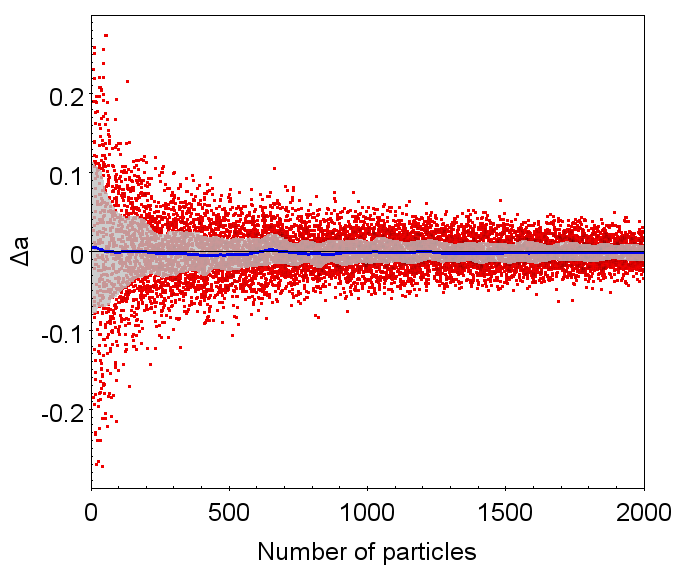}
    \includegraphics[width=5.8cm]{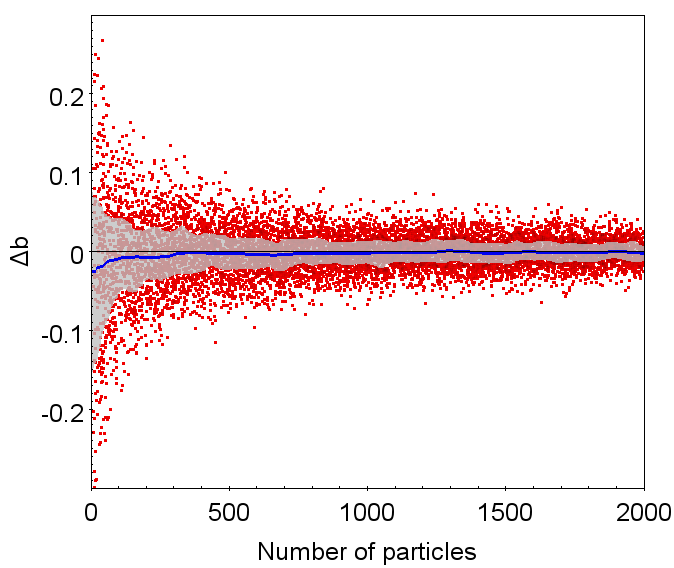}
    \includegraphics[width=5.8cm]{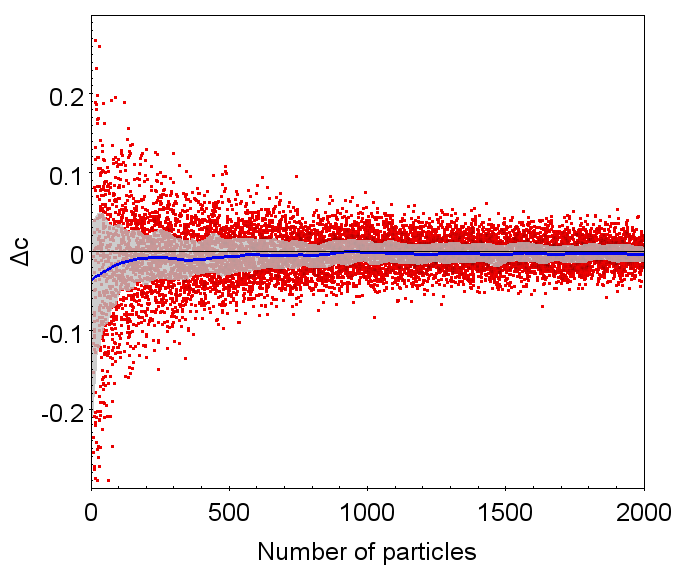}
  	\caption{Analysis of 10 000 simulated halos. Relative difference ($\Delta a$, $\Delta b$, $\Delta c$) in the recovered semi-axes (a,b,c) as a function of the number of particles of (upper panel) quasi-spherical halos with $c_{sim}/a_{sim}$=0.72 and (lower panel) flattened halos with $c_{sim}/a_{sim}$=0.25. Blue lines indicate the median of the distributions and the grey area shows the quartiles.}
    \label{simulations_axis}
\end{figure*}

\begin{figure*}
\centering
    \includegraphics[width=0.32\textwidth]{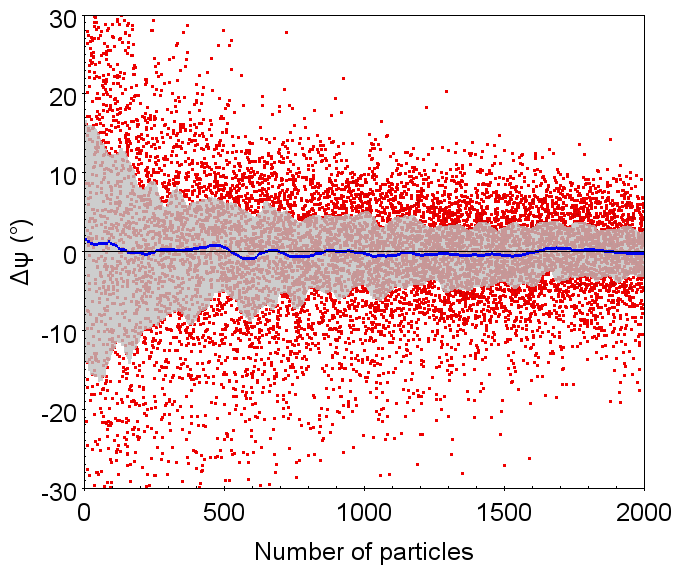}
    \includegraphics[width=0.32\textwidth]{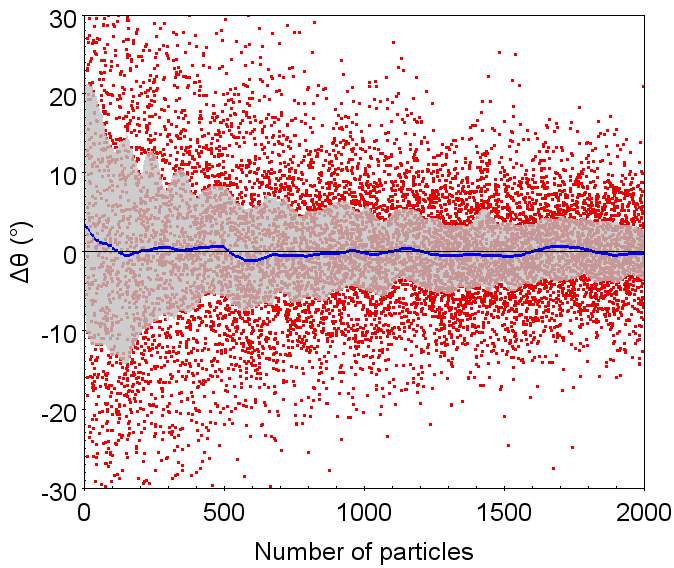}
    \includegraphics[width=0.32\textwidth]{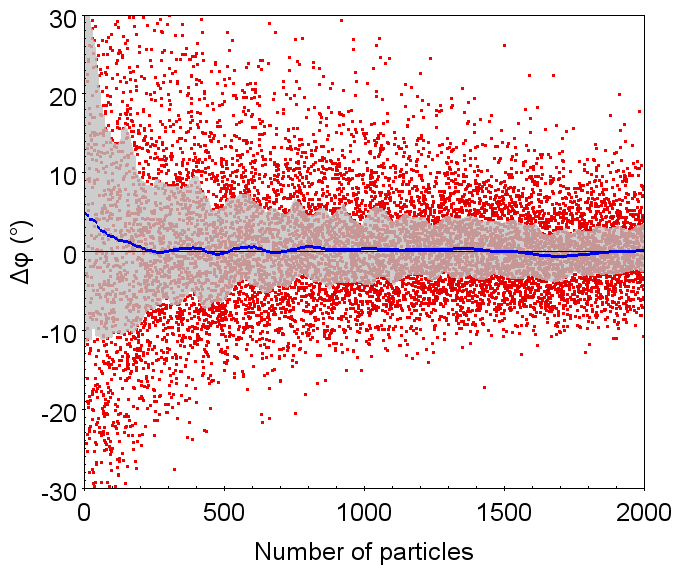}
    \includegraphics[width=0.32\textwidth]{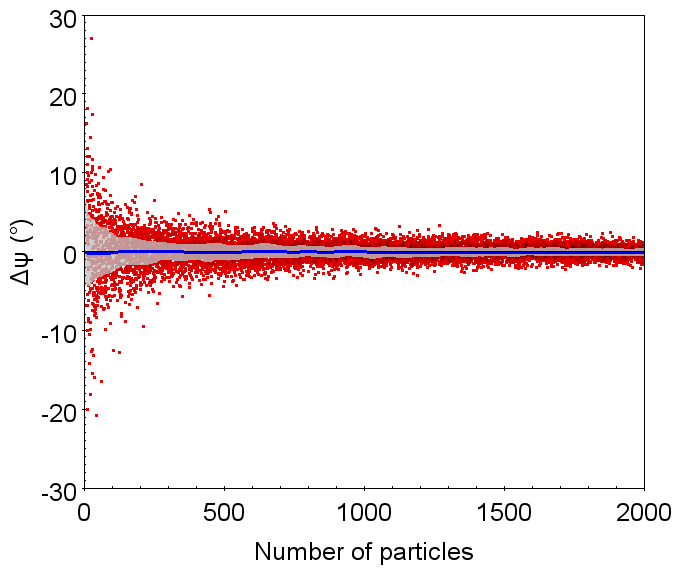}
    \includegraphics[width=0.32\textwidth]{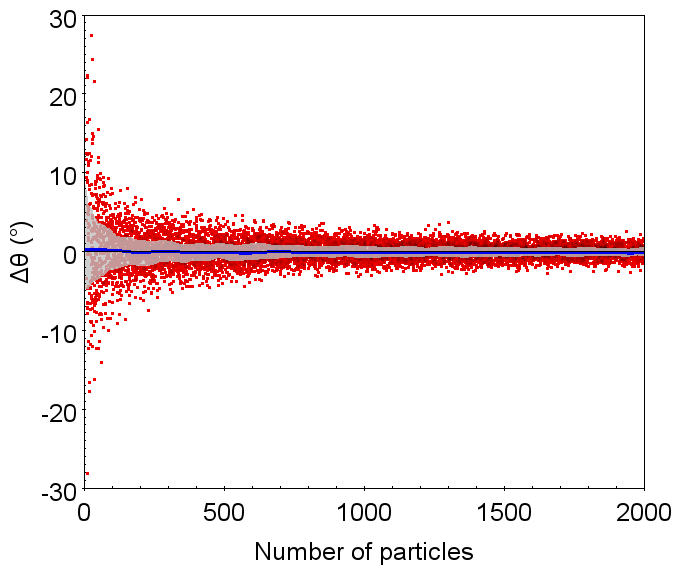} 
    \includegraphics[width=0.32\textwidth]{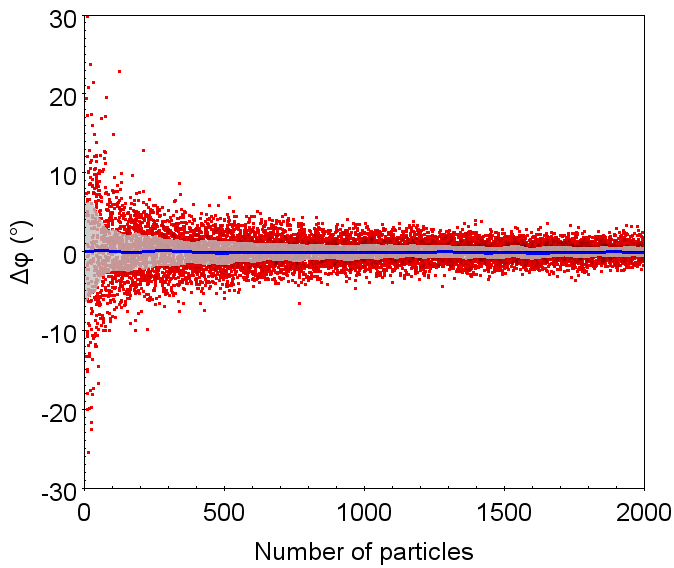}
  	\caption{Analysis of 10 000 simulated halos. Difference between the recovered angles ($\psi$, $\theta$, $\phi$) and their true simulated values expressed in deg as a function of the number of particles. Blue lines indicate the median of the distributions and the grey area shows the quartiles. Upper panel: Quasi-spherical halos with $c_{sim}/a_{sim}$=0.72. Lower panel: Flattened halos with $c_{sim}/a_{sim}$=0.25.}
    \label{simulations_angle}
\end{figure*}

\begin{figure*}
\centering
    \includegraphics[width=0.32\textwidth]{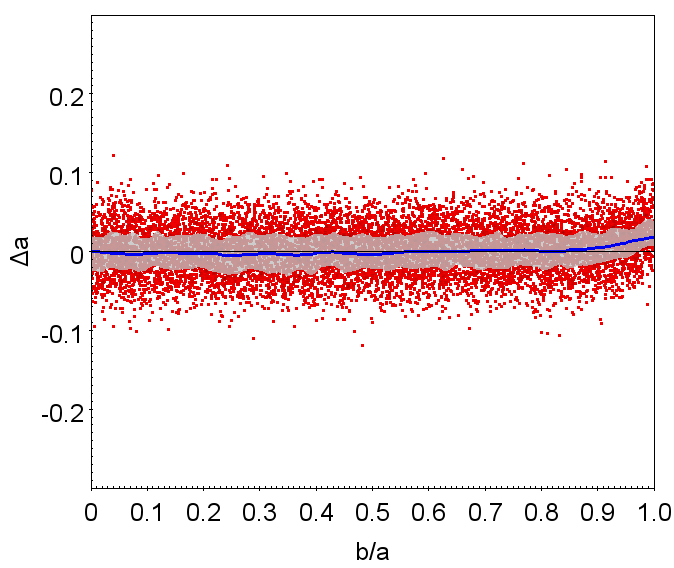}
    \includegraphics[width=0.32\textwidth]{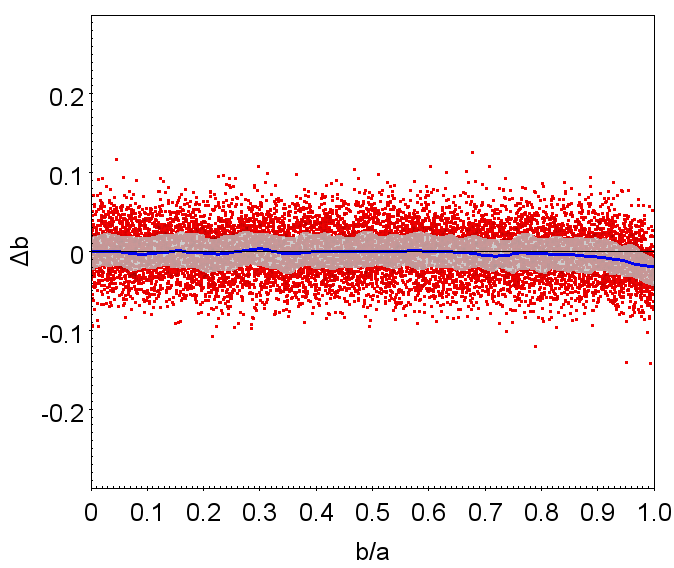} 
    \includegraphics[width=0.32\textwidth]{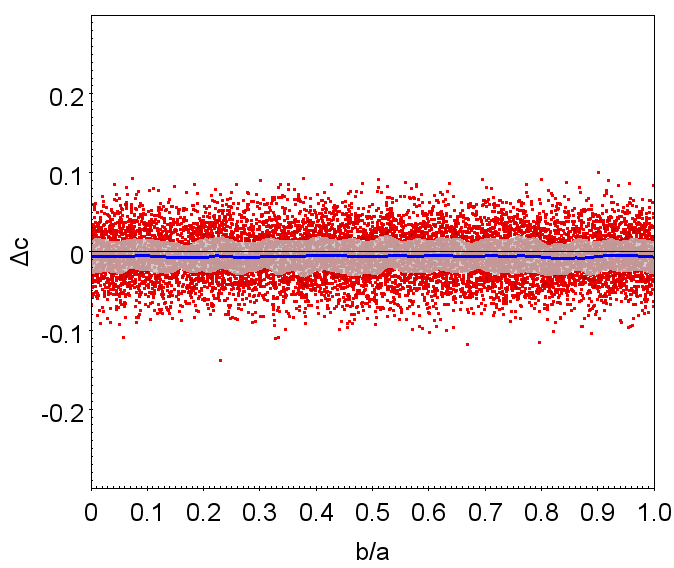}
    \label{simulations_axis_ba_axis}
    \includegraphics[width=0.32\textwidth]{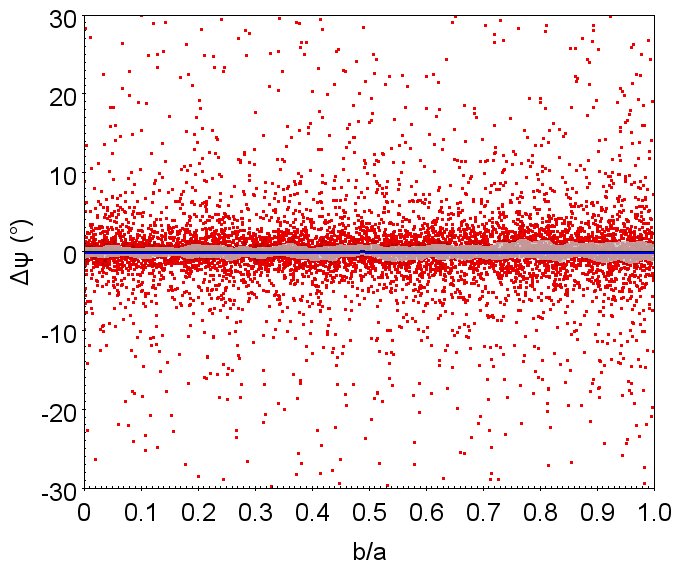}
    \includegraphics[width=0.32\textwidth]{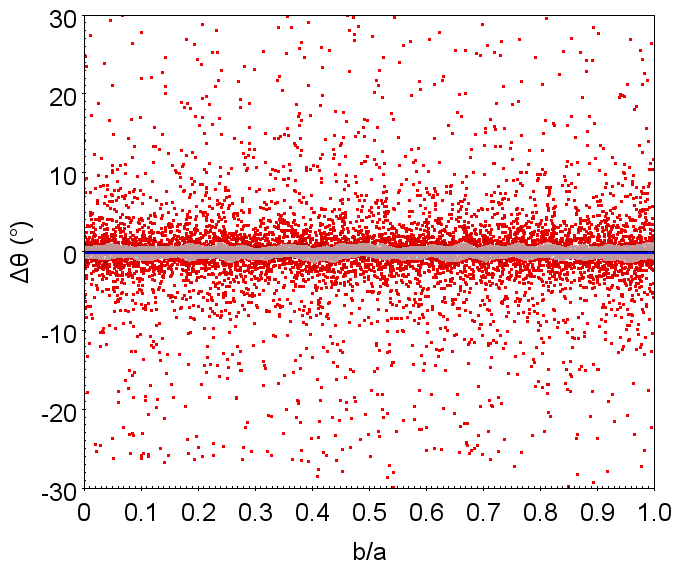} 
    \includegraphics[width=0.32\textwidth]{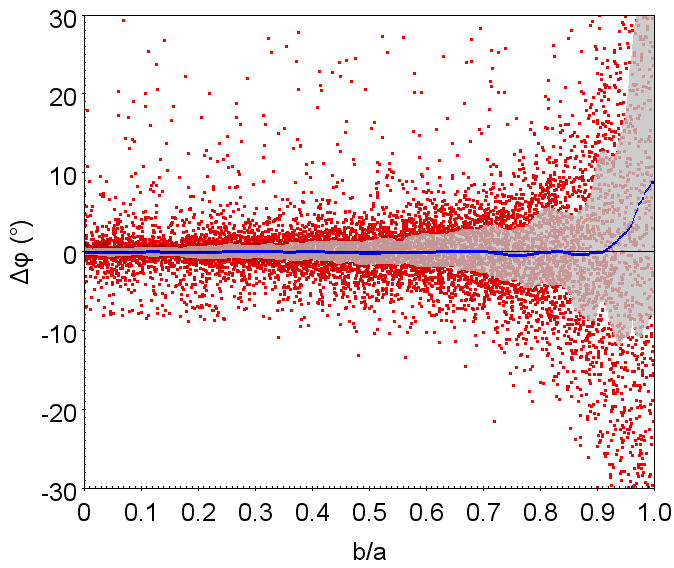}
  	\caption{Analysis of 10 000 simulated halos with 500 particles. Upper panel: Relative differences ($\Delta a$, $\Delta b$, $\Delta c$) in the recovered semi-axes (a,b,c) as a function of the sphericity (c/a). Lower panel: Difference between the recovered angles ($\psi$, $\theta$, $\phi$) and their true simulated values expressed in deg as a function of the sphericity (c/a). Blue lines indicate the median of the distributions and the grey area shows the quartiles.}     
   \label{simulations_ba}
\end{figure*}

\end{appendix}

\end{document}